\documentclass[
reprint,
superscriptaddress,
nofootinbib,
 amsmath,amssymb,
 aps,
 prb
]{revtex4-2}
\usepackage[compat=1.1.0]{tikz-feynhand}
\usepackage{amsmath,amssymb,bm,braket,url,dsfont}
\usepackage{mathtools}
\usepackage{graphicx}
\usepackage{xcolor}
\usepackage{longtable}
\usepackage{multirow}
\usepackage{array}
\usepackage{booktabs}
\usepackage{physics}
\usepackage{here}
\usepackage[version=3]{mhchem}
\usepackage{siunitx}

\usepackage[
draft=false,
bookmarks=true,
bookmarksnumbered=false,
bookmarksopen=false,
bookmarksopenlevel=4,
colorlinks=true,
anchorcolor=black,
citecolor=blue,
filecolor=magenta,
linkcolor=blue,
linkbordercolor={1 0 0},
urlcolor=blue,
pdfborder={0 0 1},
pdftitle={},
pdfauthor={},
pdfsubject={},
pdfkeywords={},
pdfpagemode=UseThumbs,
]{hyperref}

\newcommand{\expecval}[1]{\left \langle {#1} \right \rangle}
\newcommand{\figref}[1]{FIG.~\ref{#1}}
\newcommand{\tabref}[1]{TABLE~\ref{#1}}
\newcommand{\appref}[1]{Appendix~\ref{#1}}
\newcommand{\Eqref}[1]{Eq.~\eqref{#1}}

\begin{document}

\title{
        Bulk photovoltaic effect in antiferromagnet: Role of collective spin dynamics
}

\author{Junta Iguchi}
\email{iguchi-junta688@g.ecc.u-tokyo.ac.jp}
\affiliation{Department of Applied Physics, The University of Tokyo, {Bunkyo}, Tokyo 113-8656, Japan}

\author{Hikaru Watanabe} 
\email{hikaru-watanabe@g.ecc.u-tokyo.ac.jp}
\affiliation{Research Center for Advanced Science and Technology, The University of Tokyo, {Meguro}, Tokyo 153-8904, Japan}

\author{Yuta Murakami}
\affiliation{Center for Emergent Matter Science, RIKEN, {Wako}, Saitama 351-0198, Japan}

\author{Takuya Nomoto} 
\affiliation{Research Center for Advanced Science and Technology, The University of Tokyo, {Meguro}, Tokyo 153-8904, Japan}

\author{Ryotaro Arita} 
\affiliation{Research Center for Advanced Science and Technology, The University of Tokyo, {Meguro}, Tokyo 153-8904, Japan}
\affiliation{Center for Emergent Matter Science, RIKEN, {Wako}, Saitama 351-0198, Japan}

\begin{abstract}
Inspired by recent advancements in the bulk photovoltaic effect which can extend beyond the independent particle approximation (IPA), this study delves into the influence of collective spin dynamics in an antiferromagnetic on photocurrent generation using a time domain calculation. In the linear and photocurrent conductivity spectra, we observe peaks below the bandgap regime, attributed to the resonant contributions of collective modes, alongside broadband modifications resulting from off-resonant spin dynamics. Notably, the emergence of spin dynamics allows various types of photocurrent, which are absent in the IPA framework.
Furthermore, we emphasize the importance of energy scale proximity between electronic and spin degrees of freedom in enabling efficient feedback. These findings offer new avenues for efficient energy harvesting and optoelectronic applications.
\end{abstract}

\maketitle
\section{introduction}
The photovoltaic effect is a phenomenon in which light is converted into an electric current known as a photocurrent. It can be understood as an even-order optical response, hence requires inversion breaking in the system. This effect is commonly observed in heterostructures such as p-n junctions \cite{Williams1960}, where inversion symmetry is artificially broken. Recently, the growing demand for energy harvesting has provoked the exploration for photocurrent generation in bulk single-phase crystals without inversion symmetry, termed the bulk photovoltaic effect (BPVE) \cite{Fridkin2001, Baltz1981}. BPVE has attracted much attention due to its distinct features. For instance, unlike conventional PVE, photovoltage is not limited by the band gap energy, and the Shockley-Queisser limit \cite{Shockley1961} could be overcome \cite{Spanier2016}. 

Extensive studies have revealed the complete classification of BPVE in non-magnetic and magnetic systems based on the independent particle approximation (IPA) \cite{Sipe2000, Ahn2020, Watanabe2021}. For a representative example, BPVE in insulators can be classified into two contributions: shift current and injection current. The shift current \cite{Young2012, Sotome2019, Sotome2021, Tan2016} arises from the difference in intracell coordinates of excited carriers and is related to the geometric quantity called Berry connection. On the other hand, injection current results from the group velocity difference of the excited electron-hole pair. Based on this classification, extensive searches for efficient BPVE have been carried out across a wide range of materials, including Weyl semimetals \cite{Osterhoudt2019}, topological insulators \cite{Braun2016}, and transition metal dichalcogenides \cite{Yu2016}.

This classification is based on the IPA with the rigid band picture. However, recent progress in theoretical studies of BPVE has revealed that electron correlation effects in solids also play an important role in photocurrent generation \cite{Morimoto2018, Michishita2021}. In particular, photocurrent can originate from elementary excitations below the bandgap, such as excitons \cite{Morimoto2016_exciton, Chan2021}, soft phonons in ferroelectrics \cite{Okamura2022}, electromagnons in multiferroics \cite{Morimoto2019_electromagnon}, and collective modes in excitonic insulators \cite{Kaneko2021}. Understanding these phenomena requires explicit consideration of collective dynamics beyond the IPA.  

In this paper, we focus on the BPVE in an antiferromagnetic system and discuss the effects of the collective spin dynamics, exploring BPVE beyond IPA. BPVE in magnetic materials is particularly intriguing for several reasons. 
Firstly, there are various kinds of magnetic textures, in some of which the magnetic order breaks the inversion symmetry and thereby ties the charge degree of freedom with spins. This strong spin-charge coupling is also attracting much attention from the field of multiferroics \cite{Tokura2014, Spaldin2019, Fiebig2005} and antiferromagnetic spintronics \cite{Jungwirth2016, RevModPhys_Baltz, RevModPhys.91.035004, Amin2020}. 
Secondly, the spin textures accompany the breaking of time-reversal symmetry, which may allow the nonzero contribution from injection current under linearly polarized and unpolarized light. This is advantageous because the amplitude of the linear injection can dominate that of the shift currents, particularly in clean systems, potentially enabling highly efficient photovoltaic devices.
Thirdly, not limited to the antiferromagnetic system, the external magnetic field can tune the magnetic structures, and we can expect a significant modulation of photocurrent response. In fact, a recent study on the BPVE effect in magnetic systems within the IPA shows that photocurrent direction can be controlled by external field \cite{Okumura2021, Zhang2019}, demonstrating the high tunability of the photocurrent in magnets \cite{Watanabe2020}. 
As illustrated, BPVE phenomena in magnetic systems exhibit rich characteristics, even within the IPA framework. In such systems, the role of spin dynamics on BPVE remains largely unexplored. Especially in antiferromagnetic systems, the energy scale of the collective spin dynamics typically lies in the THz regime, which is proximate to that of the electronic system. Hence, we expect that the spin dynamics significantly affects the response of the electronic systems. Moreover, the emergence of fictitious fields by the spin dynamics may lead to richer optoelectronic properties, which are absent within IPA \cite{Kaneko2021}. By investigating the effects of spin dynamics on PVE, we aim to provide a more comprehensive understanding of BPVE, moving beyond the IPA. 

In this paper, we investigate the role of collective spin dynamics in BPVE in a simple model representing a one-dimensional antiferromagnet. We employ a unified framework for a real-time simulation of conduction electrons and localized spin moments based on the methods introduced in previous studies \cite{Kaneko2021, Ono2021}. This approach allows us to investigate the influence of collective modes on BPVE and gives a comprehensive understanding of their impact on the optical response. 
By employing the symmetry analysis, we identify an optically active collective mode and investigate its role in both linear and nonlinear optical responses. In the conductivity spectrum, we observe a peak below the bandgap regime stemming from resonant contributions of collective modes and broadband modifications resulting from off-resonant contributions of collective spin dynamics.
Additionally, we found that photocurrent arising from the spin dynamics can be classified into several types, which are absent in the IPA. Here, the symmetry and the phase degrees of freedom of the driven forces play crucial roles. 
Furthermore, we demonstrate the significance of proximity in energy scales between collective modes and electronic excitations. This insight provides a guiding principle for harnessing substantial photocurrents from spin dynamics.
\section{method}
In this section, we elaborate on the model and explain how we incorporate the effect of collective spin dynamics on optical responses. Firstly, we introduce the model used in this study, then move on to our time-dependent simulation scheme.
\subsection{Model}
\begin{figure}
        \centering
        \includegraphics[width=\linewidth]{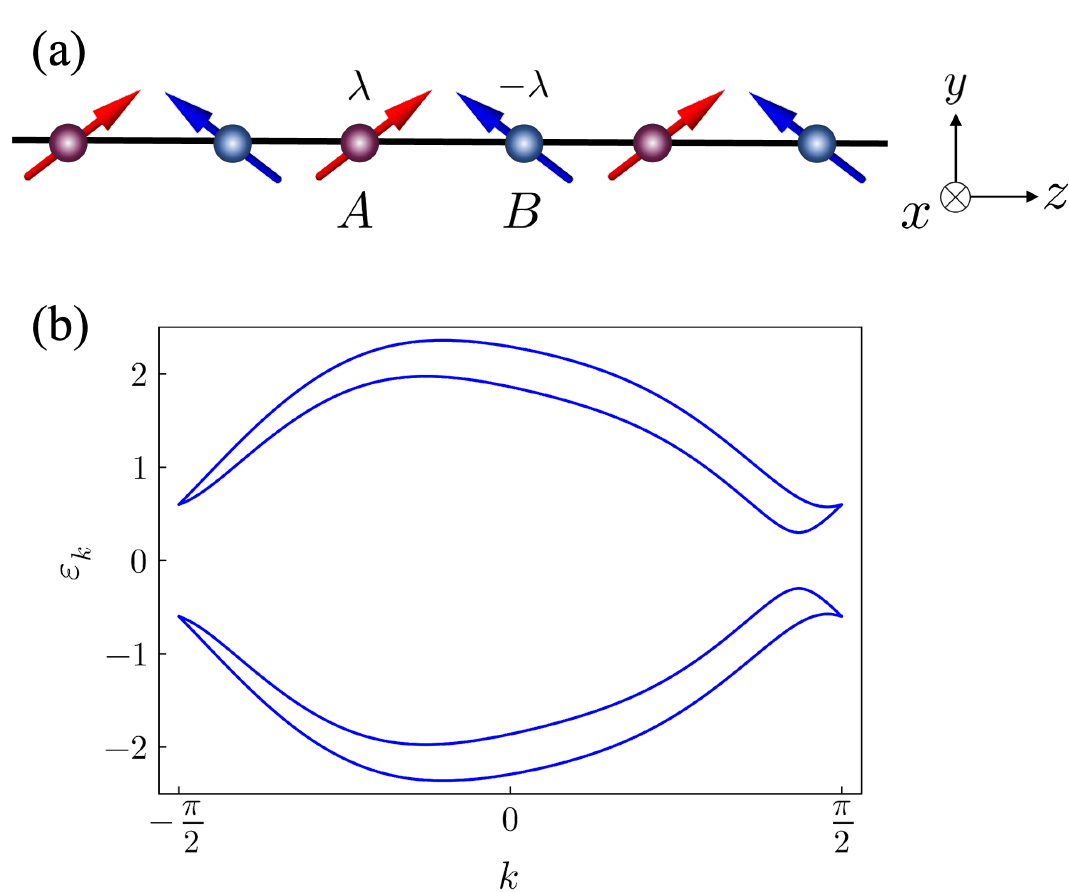}
        \caption{(a) One-dimensional chain model with canted antiferromagnetic order. (b) band dispersion of the system with the parameters $t_{h} = 1, \lambda = 0.8, J = 0.6, K_{z} = 0.2, h_{y} = 0.2$.}
        \label{model_pic}
\end{figure}
This study focuses on a locally noncentrosymmetric system in one dimension, where conduction electrons are coupled to localized spin moments with a canted antiferromagnetic structure (see \figref{model_pic}).
This model is a simple representative example of magnetoelectric materials \cite{Yanase2014}, where we expect large spin and charge coupling.

The Hamiltonian of the model is expressed as
\begin{align}
        \hat{\mathcal{H}} = \hat{\mathcal{H}}_{\mathrm{ele}} + \hat{\mathcal{H}}_{\mathrm{exc}} + \mathcal{H}_{\mathrm{spin}} + \hat{\mathcal{H}}_{E}.
\end{align}
The first term,
\begin{align}
        \hat{\mathcal{H}}_{\mathrm{ele}} = -2t_{h}\sum_{k}\sum_{\sigma}\cos k\qty[\hat{c}_{A\sigma}^{\dagger}(k)\hat{c}_{B\sigma}(k) + \hat{c}_{B\sigma}^{\dagger}(k)\hat{c}_{A\sigma}(k)] \nonumber\\ 
        -\lambda\sum_{k}\sum_{\sigma\sigma^{\prime}}\qty(\vb{g}(k)\cdot \vb*{\sigma})^{\sigma\sigma^{\prime}}\qty[\hat{c}_{A\sigma}^{\dagger}(k)\hat{c}_{A\sigma^{\prime}}(k) - \hat{c}_{B\sigma}^{\dagger}(k)\hat{c}_{B\sigma^{\prime}}(k)]
\end{align}
is the Hamiltonian of the electronic system, consisting of the nearest neighbor hopping $t_{h}$, and the sublattice dependent antisymmetric spin-orbit coupling $\lambda$ (sASOC). $\hat{c}_{\alpha\sigma}^{\dagger}(k)$ ($\hat{c}_{\alpha\sigma}(k)$) is the creation (annihilation) operator of the electron on sublattice $\alpha$ ($\alpha = A, B$) having spin $\sigma$ ($\sigma=\uparrow, \downarrow$), and $k$ is the wave vector.
$\vb{g}(k)$ is called g-vector, and satisfies $\vb{g}(-k) = -\vb{g}(k)$. We set $\vb{g}(k) = (0 ,0, \sin 2k)$, implicitly assuming the presence of ligand structure around each sublattice which breaks the local inversion symmetry. This sASOC is the essential ingredient to enhance magnetoelectric coupling \cite{Yanase2014, Zelenzny2014, Hayami2014}.

The second term
\begin{align}
        \hat{\mathcal{H}}_{\mathrm{exc}} = -J\sum_{k}\sum_{\alpha}\sum_{\sigma\sigma^{\prime}}\hat{c}_{\alpha\sigma}^{\dagger}(k)\qty(\vb*{\sigma}\cdot\vb{S}_{\alpha}(t))^{\sigma\sigma^{\prime}}\hat{c}_{\alpha\sigma^{\prime}}(k)
        \label{Hund_Hamiltonian}
\end{align}
corresponds to the interaction between electronic degrees of freedom and localized spin moments with coupling constant $J$. Here we assume that the localized spins $\mathrm{S}_{\alpha}$ are classical spins and set $\abs{\vb{S}_{\alpha}}=1$. We note that, in \Eqref{Hund_Hamiltonian}, we assume that the spin configuration is commensurate with the lattice structure as in \figref{model_pic} and the local spins can be classified only by the sublattices.

The third term 
\begin{align}
        \mathcal{H}_{\mathrm{spin}} = -\sum_{\alpha} \qty[ K_{z}\qty(\mathrm{S}_{\alpha}^{z})^{2} - h_{y}\mathrm{S}_{\alpha}^{y} ]
\end{align}
is the Hamiltonian for the localized spins. To stabilize the antiferromagnetic order along the chain direction ($z$ direction), we consider the uniaxial anisotropy $K_{z}$. We also apply the external magnetic field perpendicular to the chain direction ($y$ direction) to induce the canted configurations. This is because, without this canted moment, the light-induced spin accumulation on each sublattice, which acts as spin transfer torque for the localized spin moments, would be parallel to $\vb{S}_{\alpha}$. Therefore, there would be no spin dynamics induced by light irradiation.

The last term 
\begin{align}
        \hat{\mathcal{H}}_{E} = -E^{z}(t)\sum_{kk^{\prime}}\sum_{\alpha}\sum_{\sigma}\qty[i\pdv{k}\delta(k-k^{\prime})]\hat{c}_{k\alpha\sigma}^{\dagger}\hat{c}_{k^{\prime}\alpha\sigma} \label{light-matter-coupling}
\end{align}
describes the light-matter coupling in the length gauge, where $E^{z}(t)$ is a time-dependent light field along the chain direction, where we set lattice constant $a = 1$, and elementary charge $e=1$.
In the expression \Eqref{light-matter-coupling}, we assume that the Wannier state of a conducting electron is well-localized at a given site and ignore the light-matter coupling originating from its spatial extension. This light-matter coupling is the same as the celebrated Peierls substitution in the dipolar gauge \cite{Murakami2022}. The difficulty of dealing with the $k$-derivative of the delta function in \Eqref{light-matter-coupling} can be resolved in the following subsection.
\subsection{Calculation Scheme}
Here, we introduce two coupled equations that describe the dynamics of electronic degrees of freedom and localized spin moment. 
Firstly, the time evolution of the conduction electrons can be described by the single-particle density matrix (SPDM) $\rho_{\alpha \beta}^{\sigma \sigma^{\prime}}(k) = \expecval{\hat{c}_{\beta\sigma^{\prime}}^{\dagger}(k)\hat{c}_{\alpha\sigma}(k)}$, and SPDM satisfies the following equation called the von Neumann equation \cite{Yue2022},
\begin{align}
        \begin{split}
            \pdv{\vb*{\rho}(k, t)}{t} = -i\qty[\vb*{H}(k,t),\, \vb*{\rho}(k, t)] - E^{z}(t)\pdv{\vb*{\rho}(k, t)}{k} \\
        - \gamma(\vb*{\rho}(k, t) - \vb*{\rho}_{\text{eq}}(k)).
        \label{vonNeumann}
        \end{split}
\end{align}
Secondly, the time evolution of the localized spin system is described by the Landau-Lifshitz-Gilbert (LLG) equation. 
\begin{align}
        \dfrac{d\boldsymbol{\mathrm{S}}_{\alpha}}{dt} &= \dfrac{1}{1 + \alpha_{G}^{2}}\left( \boldsymbol{\mathrm{H}}_{\alpha}^{\mathrm{eff}} \times \boldsymbol{\mathrm{S}}_{\alpha} + \alpha_{G} \boldsymbol{\mathrm{S}}_{\alpha} \times \left( \boldsymbol{\mathrm{S}}_{\alpha} \times\boldsymbol{\mathrm{H}}_{\alpha}^{\mathrm{eff}} \right)\right) \label{LLG}, \\
        \boldsymbol{\mathrm{H}}_{\alpha}^{\mathrm{eff}} &= -J\langle \boldsymbol{\sigma}_{\alpha} \rangle + \dfrac{\delta \mathcal{H}_{\text{spin}}}{\delta \boldsymbol{\mathrm{S}}_{\alpha}}.
\end{align}
In \Eqref{vonNeumann}$, \vb*{H}(k,t)$ is the time-dependent electronic Hamiltonian at each $k$ point defined as follows
\begin{align}
        \hat{\mathcal{H}}_{\mathrm{ele}} + \hat{\mathcal{H}}_{\mathrm{exc}} = \sum_{k}\sum_{\alpha\beta}\sum_{\sigma\sigma^{\prime}}\qty(\vb*{H}(k,t))_{\alpha\beta}^{\sigma\sigma^{\prime}}\hat{c}_{\alpha\sigma}^{\dagger}(k)\hat{c}_{\beta\sigma^{\prime}}(k). \label{time_dependent_hamiltonian}
\end{align}
The $k$-derivative of the delta function in \Eqref{light-matter-coupling} becomes the $k$-derivative of the SPDM, which is computationally manageable.
In the LLG equation \Eqref{LLG}, $\expecval{\vb*{\sigma}_{\alpha}}$ is the sublattice-dependent spin density of itinerant electrons, and this can be calculated from SPDM. This method allows us to capture both the dynamics of electronic and spin systems in the time domain.

Additionally, real materials exhibit relaxation of excited carriers due to electron-electron correlations, electron-phonon interactions, and impurity scattering. To ensure a physically reasonable response to light, we account for these effects phenomenologically by using the relaxation time approximation in the von Neumann equation as $\gamma(\vb*{\rho}(k, t) - \vb*{\rho}_{\text{eq}}(k))$ in \Eqref{vonNeumann}, and the Gilbert damping $\alpha_{G}$ in \Eqref{LLG}. Here $\vb*{\rho}_{\text{eq}}(k)$ is the SPDM in the equilibrium at the temperature $T=0$ as shown below.

The SPDM at equilibrium $\vb*{\rho}_{\text{eq}}(k)$ represents the SPDM in the initial state at temperature $T=0$. SPDM in the band basis $\tilde{\vb*{{\rho}}}_{\text{eq}}(k)$ is diagonal, and the diagonal component represents the occupation number $\Theta(\mu - \epsilon_{nk})$, with $\epsilon_{nk}$ being the eigenvalue of the Hamiltonian $\vb*{H}(k)$. Therefore $\tilde{\vb*{\rho}}_{\text{eq}}(k)$ is written as,
\begin{align}
        \qty(\tilde{\vb*{{\rho}}}_{\text{eq}}(k))_{nn^{\prime}} = \delta_{nn^{\prime}}\Theta(\mu - \epsilon_{nk}).
\end{align}
By using this expression, the SPDM in the original basis $\vb*{\rho}_{\text{eq}}(k)$ can be calculated by 
\begin{align}
        \vb*{\rho}_{\text{eq}}(k) = \vb*{U}(k)\tilde{\vb*{\rho}}_{\text{eq}}(k)\vb*{U}^{\dagger}(k)
\end{align}
with $\vb*{U}$ is a unitary matrix which diagonalizes the Hamiltonian as, 
\begin{align}
        \vb*{U}^{\dagger}(k)\vb*{H}(k)\vb*{U}(k) = \vb*{\mathcal{E}}(k),\\
        \qty(\vb*{\mathcal{E}}(k))_{nn^{\prime}} = \delta_{nn^{\prime}}\epsilon_{nk}.
\end{align}
To obtain initial spin configuration, where the localized spin moment $\vb{S}_{\alpha}(t=0)$ is parallel to the effective field $\vb{H}_{\alpha}^{\mathrm{eff}}$ acting on $\vb{S}_{\alpha}(t = 0)$, we perform the self-consistent calculation.

In each time step, the current density is evaluated from SPDM as follows. 
The current operator in the length gauge is expressed as 
\begin{align}
        \hat{J}(t) &= \sum_{k}\sum_{\alpha\beta}\sum_{\sigma\sigma^{\prime}}\pdv{\qty(\vb*{H}(k,t))_{\alpha\beta}^{\sigma\sigma^{\prime}}}{k}\hat{c}_{\alpha\sigma}^{\dagger}(k)\hat{c}_{\beta\sigma^{\prime}}(k) \\
        &\equiv \sum_{k}\sum_{\alpha\beta}\sum_{\sigma\sigma^{\prime}}\qty(\vb*{J}(k))_{\alpha\beta}^{\sigma\sigma^{\prime}}\hat{c}_{\alpha\sigma}^{\dagger}(k)\hat{c}_{\beta\sigma^{\prime}}(k).
\end{align}
It is noteworthy that since the $\hat{\mathcal{H}}_{\text{exc}}$ term in the Hamiltonian does not depend on momentum $k$, the current operator is independent of local spin dynamics.

We solve the coupled equations \Eqref{vonNeumann} and \Eqref{LLG}  by the fourth-order Runge-Kutta method. In each time step, we calculate the sublattice-dependent spin density using the SPDM and update the exchange Hamiltonian with the newly obtained spin configurations. Additionally, the expectation value of the current operator is obtained by using the density matrix as 
\begin{align}
        J^{z}(t) = \sum_{k}\operatorname{Tr}\qty[\vb*{J}(k)\vb*{\rho}(k, t)].
\end{align}
To evaluate the $\partial\vb*{\rho}(k, t)/\partial k$ in von Neumann equation \Eqref{vonNeumann}, we use symmetric derivative
\begin{align}
        \pdv{\vb*{\rho}(k, t)}{k} = \dfrac{\vb*{\rho}(k+dk, t) - \vb*{\rho}(k-dk, t)}{2dk},
\end{align}
where $dk = \pi/N$, with $N = 5000$.

Based on this scheme, we calculate the linear and nonlinear susceptibility to the light field from the obtained current response, which will be shown in the next section. In the following calculations, we use the parameters $t_{h} = 1, \lambda = 0.8, J = 0.6, K_{z} = 0.2, h_{y} = 0.2, \alpha_{G} = 0.01, \gamma = 0.01$, otherwise explicitly mentioned. In \figref{model_pic} (b), we show the band structure of the system in equilibrium. Due to the inversion breaking and time reversal breaking in the system, the band structure is asymmetric about $k$, namely $\epsilon(k) \neq \epsilon(-k)$. We set the chemical potential $\mu$ to be $0$ within the bandgap. 

\section{Results}
This section presents the result of the light-induced dynamics of the system. Starting from the symmetry analysis, we clarify the optically active collective excitations in \ref{symmetry_analysis}. Secondly, in \ref{linear_response_functions}, we investigate the effect of spin dynamics on optical response in the linear response regime. In \ref{photocurrent_spectra} and \ref{decomposition_of_photocurrent}, we study the effect of collective mode on photocurrent conductivity. Here, we decompose the photocurrent into different processes, where the linear susceptibility of spins to light field obtained in  \ref{linear_response_functions} plays a crucial role in photocurrent generation. The last subsection \ref{Tuning_of_photocurrent} is dedicated to demonstrating the tunability of photocurrent generation by changing the canted angle of the local spin moments. 
\subsection{Symmetry analysis}\label{symmetry_analysis}
\begin{figure}
        \centering
        \includegraphics[width=\linewidth]{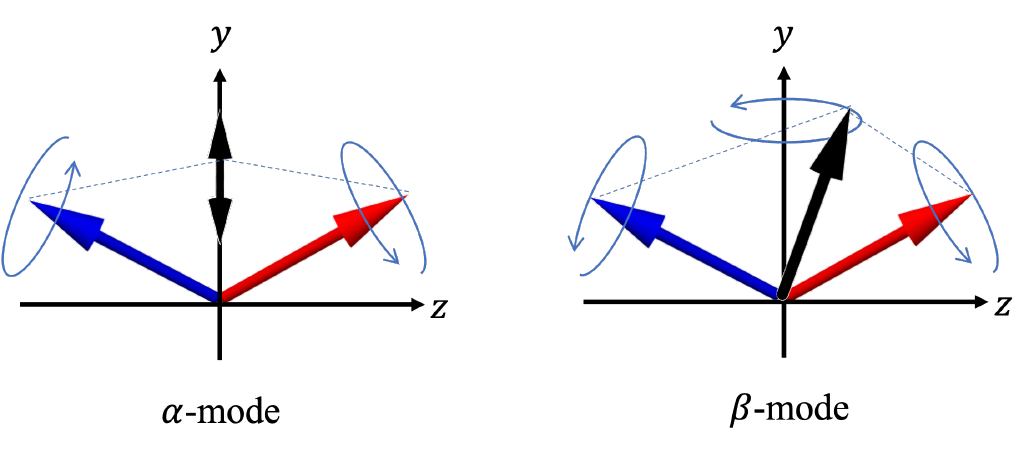}
        \caption{Collecive mode of canted antiferromagnetic moment. The red (blue) arrow indicates the spin moment in the $A$ ($B$) sublattice, respectively. The black arrow denotes the summation of the sublattice spin moment.}
        \label{Col_mode}
\end{figure}

\begin{table}
        \caption{Sign under symmetry operation of magnetic point group $\mathcal{G}$. $\mathrm{M}^{a}$ means ferroic configurations of spin moment along the $a$ direction, while $\mathrm{L}^{a}$ means staggered spin moments along $a$ direction. $+$ means the observable does not change its sign, while $-$ means the observable flip its sign under the operation.}
        \begin{ruledtabular}
        \begin{tabular}{ccccc}
                & $1$ & $\theta2_{x}$ & $\bar{2}_{y}$ & $\theta \bar{2}_{z}$ \\\colrule
                $E^{z}$ & $+$ & $-$           & $+$       & $-$ \\
                $J^{z}$ & $+$ & $+$           & $+$       & $+$  \\
        $\mathrm{M}^{x}$& $+$ & $-$           & $-$       & $+$  \\
        $\mathrm{L}^{x}$& $+$ & $-$           & $+$       & $-$ \\
        $\mathrm{M}^{y}$& $+$ & $+$           & $+$       & $+$ \\
        $\mathrm{L}^{y}$& $+$ & $+$           & $-$       & $-$ \\
        $\mathrm{M}^{z}$& $+$ & $+$           & $-$       & $-$ \\
        $\mathrm{L}^{z}$& $+$ & $+$           & $+$       & $+$ \\
        \end{tabular}
        \label{tab:symmetry}
        \end{ruledtabular}
\end{table}

In this subsection, we analyze the collective spin dynamics that is linearly coupled to the external light field based on the symmetry analysis. Since photocurrent response is largely restricted by the symmetry of the system and the external field, symmetry analysis of collective dynamics lays the foundation of our study.

The system has the following symmetries and belongs to the magnetic point group $\mathcal{G} = 2^{\prime}mm^{\prime}$, explicitly given by
\begin{align}
        \mathcal{G} = \qty{1, \theta2_{x}, \bar{2}_{y}, \theta \bar{2}_{z}}.
\end{align}
Here, $1, \theta, 2_{a}$, and $\bar{2}_{a}$ are the identity operator, time reversal operator, twofold rotation around the $a$ axis, rotatory inversion about the $a$ axis, respectively. 

Let us identify the collective modes in the canted antiferromagnetic moments. Since our method is based on the momentum-space formulation, we only consider $k=0$ magnon, or antiferromagnetic resonance. As we show in  \figref{Col_mode}, two collective modes exist in our system, named $\alpha$-mode and $\beta$-mode \cite{Rezende2019}. In the $\alpha$-mode, the total spin moment oscillates along the $y$-direction, whereas the $\beta$-mode denotes the precession of net spin moment around the $y$-axis. 

To determine which mode is optically active, we analyze the sign of observables under the symmetry operation of $\mathcal{G}$. For example, the $\bar{2}_{y}$ operation can be explicitly expressed as
\begin{align}
        \bar{2}_{y} = \qty(-i\sigma_{y}) \otimes \tau_{x}, 
\end{align}
where $\sigma$ and $\tau$ are Pauli matrices representing the spin and sublattice degrees of freedom.
Under this operation, staggered spin moment along the $x$ direction $\sigma_{x} \otimes \tau_{z}$ can be transformed as
\begin{align}
        \bar{2}_{y} \qty(\sigma_{x} \otimes \tau_{z}) \bar{2}_{y}^{-1} = \sigma_{x}\otimes\tau_{z}.
\end{align}
Therefore, $\sigma_{x}\otimes\tau_{z}$ does not change its sign under the $\bar{2}_{y}$ operation. In \tabref{tab:symmetry}, we summarize how the physical quantities change their sign under the operation in $\mathcal{G}$. Here $E^{z}, J^{z}$ are light field and electric current along the $z$ direction, respectively. $\mathrm{M}^{a}$ means ferroic configurations of spin moment along the $a$ direction, while $\mathrm{L}^{a}$ means staggered spin moments along the $a$ direction.
Owing to the incompatibility of the external light field $E^{z}$ and $\mathrm{M}^{x}, \mathrm{L}^{y}, \mathrm{M}^{z}$, under $\bar{2}_{y}$ operation, the $\beta$-mode is not linearly coupled to the light field. Although the symmetry of $\mathrm{M}^{y}, \mathrm{L}^{z}$ and $E^{z}$ are different under $\theta2_{x}, \theta \bar{2}_{z}$ operation, these can be linearly excited by electric current $J^{z}$. This is because, in nonequilibrium phenomena, time-reversal symmetry is effectively broken by the dissipation process.
This can be viewed as an extension of the frequency response of the magnetoelectric effect and Edelstein effect \cite{Watanabe2017, Hayami2018}. Consequently, only components related to $\mathrm{L}^{x}$, $\mathrm{M}^{y}$, and $\mathrm{L}^{z}$ can be linearly excited by light, indicating that the $\alpha$-mode can be linearly coupled to the light field. Since the symmetry argument does not depend on frequency profile, it is valid for both the resonant and off-resonant conditions.

There is no symmetry operation inverting the electric current $J^{z}$ since the model shows the polar symmetry along the $z$ axis similar to that of the typical photodiodes. Notably, the electric current changes its sign under the time reversal operation, coinciding with the symmetry of the toroidal moment. At the same time, the electric polarization is forbidden to be in the chain direction due to the anti-unitary symmetry. This unique combination of magnetic and polar symmetries leads to the absence of shift current and the presence of injection current within the IPA, as proved in the \appref{sym_analysis_photocurrent}.

\subsection{Linear Response Functions}\label{linear_response_functions}
\begin{figure}
        \centering
        \includegraphics[width=\linewidth]{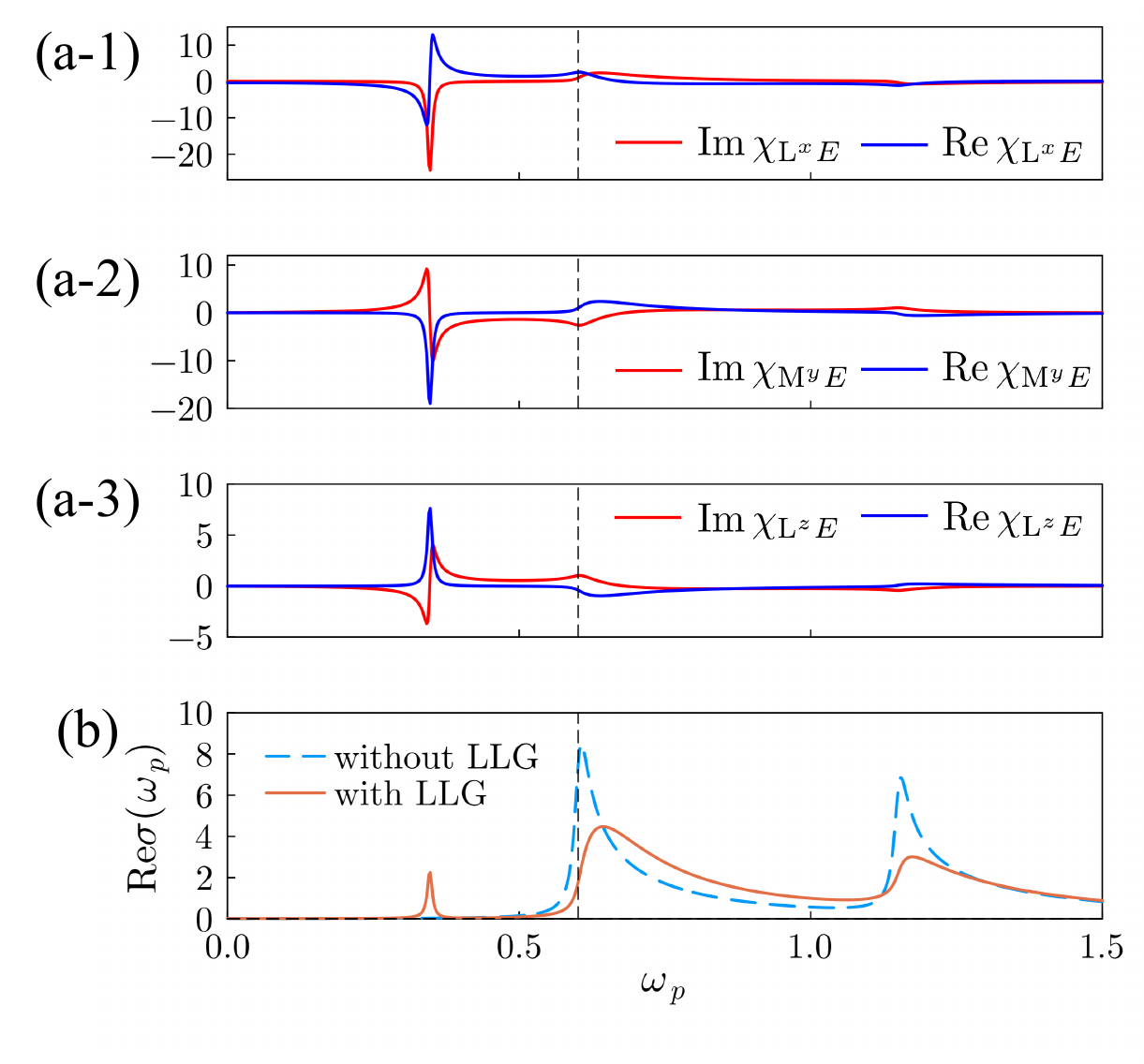}
        \caption{(a) Linear electromagnetic susceptibility to the external light field. 
        (b) Linear optical conductivity of the system. The red
        solid line and the blue dashed line indicate the calculation
        with and without updating the spin configurations, respectively. The black dashed line indicates the band gap frequency. }
        \label{linear_response_function}
\end{figure}
Before we move on to the nonlinear optical response, we examine the optical response in the linear response regime, which provides valuable insights into the mutual interaction between charge and spin degrees of freedom.
Here, we investigate the two types of linear response functions. The first one is the electromagnetic susceptibility, which has direct information on the spin collective mode induced by light. The second one is the linear optical conductivity of the system, by which we can understand how the collective excitation of the local spin affects the optical response of the electronic systems. 
Electromagnetic susceptibility is defined as
\begin{align}
        \chi_{\text{M}^{a}E}(\omega)&= \dfrac{\Delta \mathrm{M}^{a}(\omega)}{E^{z}(\omega)}, \\
        \chi_{\text{L}^{a}E}(\omega)&=\dfrac{\Delta \mathrm{L}^{a}(\omega)}{E^{z}(\omega)}.
\end{align}
Here, $\Delta \mathrm{M}^{a}(\omega), \Delta \mathrm{L}^{a}(\omega)$ are the Fourier components of the deviation of the spin configurations from the equilibrium states, which are defined as
\begin{align}
        \Delta \mathrm{M}^{a}(t) &= \frac{1}{2}\qty(\Delta \mathrm{S}_{A}^{a}(t) + \Delta \mathrm{S}_{B}^{a}(t)), \\
        \Delta \mathrm{L}^{a}(t) &= \frac{1}{2}\qty(\Delta \mathrm{S}_{A}^{a}(t) - \Delta \mathrm{S}_{B}^{a}(t)).
\end{align}
We define the spin dynamics and electromagnetic susceptibilities related to the $\alpha$-mode component as 
\begin{align}
        \Delta\vb{S}(\omega) &= \qty(\Delta\mathrm{L}^{x}(\omega), \Delta\mathrm{M}^{y}(\omega), \Delta\mathrm{L}^{z}(\omega))^{T}, \label{alpha_mode_dynamics}\\
        \vb*{\chi}_{\vb{S}E}(\omega) &= \qty(\chi_{\mathrm{L}^{x}E}(\omega), \chi_{\mathrm{M}^{y}E}(\omega), \chi_{\mathrm{L}^{z}E}(\omega))^{T} \label{electromagnetic_susceptibilities}.
\end{align}
By using these quantities, spin dynamics linearly coupled to the light field are characterized by
\begin{align}
        \Delta\vb{S}(\omega) = \vb*{\chi}_{\vb{S}E}(\omega)E^{z}(\omega).
\end{align}
Additionally, linear conductivity is defined as 
\begin{align}
        J^{z}(\omega) = \sigma(\omega)E^{z}(\omega).
\end{align}
Here, $J^{z}(\omega)$ is the Fourier transform of the $J^{z}(t)$.

To calculate these response functions, we applied a Gaussian pulse light field
\begin{align}
        E^{z}(t) = \dfrac{E_{0}}{\sqrt{2\pi\sigma^{2}}} \exp(-\dfrac{(t-t_{0})^{2}}{2\sigma^{2}})
        \label{gaussianpulse}.
\end{align}
We choose $t_{0}$ so that $E(t)\sim 0$ at $t\leq 0$. In the limit of $\sigma \to 0$, the Gaussian pulse \Eqref{gaussianpulse} becomes $E^{z}(t) = E_{0}\delta(t)$, containing all the frequencies. The linear response functions can be calculated by the Fourier transform of the resulting response in the time domain. For example, the linear optical conductivity is obtained as 
\begin{align}
        \sigma(\omega) = \dfrac{1}{E_{0}}e^{\sigma^{2}\omega^{2}/2}e^{i\omega t_{0}}\int_{0}^{\infty}e^{i\omega t}J^{z}(t)dt .
\end{align}

We plot the frequncy dependence of the linear response functions $\vb*{\chi}_{\vb{S}E}(\omega)$, and $\sigma(\omega)$ in \figref{linear_response_function}. 
In \figref{linear_response_function} (a-1)-(a-3), we show the electromagnetic susceptibility $\chi_{\vb{S}E}(\omega)$. Electromagnetic susceptibility exhibits the resonance structure below the optical gap of the electronic system. This resonant structure corresponds to the optically active $\alpha$-mode. Additionally, there are off-resonant contributions in the above bandgap regime, coming from electronic excitation.
Moreover, in the higher frequency around $\omega \sim 1.5$, $\vb*{\chi}_{\vb{S}E}$ shows small amplitude, indicating that the local spins are unaffected by the light field. This is because the local spin moments with small resonance frequency cannot follow the fast oscillations by the external field.
In the linear response calculations, we confirmed that only the $\alpha$-mode is linearly coupled to the light field, in agreement with the symmetry analysis.

In \figref{linear_response_function}(b), we show the linear optical conductivity $\operatorname{Re}\sigma^{zz}(\omega)$. The blue dashed line, denoted as without LLG, shows the linear conductivity spectra without updating the spin configuration, which corresponds to the IPA. The red solid line, denoted as with LLG, indicates the conductivity spectrum with the effect of spin dynamics.
There are three characteristics of the effect of spin dynamics. Firstly, $\operatorname{Re}\sigma(\omega)$ with LLG simulation shows the resonant peak in the in-gap regime, which is absent in the IPA. The frequency of this peak corresponds to that of the $\alpha$-mode we saw in the electromagnetic susceptibility. Secondly, the presence of spin dynamics introduces a suppression of optical conductivity within the regime above the band gap. This suppression arises due to off-resonant contribution from collective spin dynamics. It is noteworthy that due to the sum rule $\int\operatorname{Re}\sigma(\omega)d\omega = \mathrm{const.}$, the emergence of the in-gap peak results in the suppression of conductivity in the above-gap regime. 
Lastly, deviations from the IPA calculation become negligible in the higher frequency region around $\omega \sim 1.5$, as the $\operatorname{Im}\vb*{\chi}_{\mathrm{S}E}$ lose its amplitude. 

\subsection{Photocurrent Spectra}\label{photocurrent_spectra}
\begin{figure*}
        \centering
        \includegraphics[width=\linewidth]{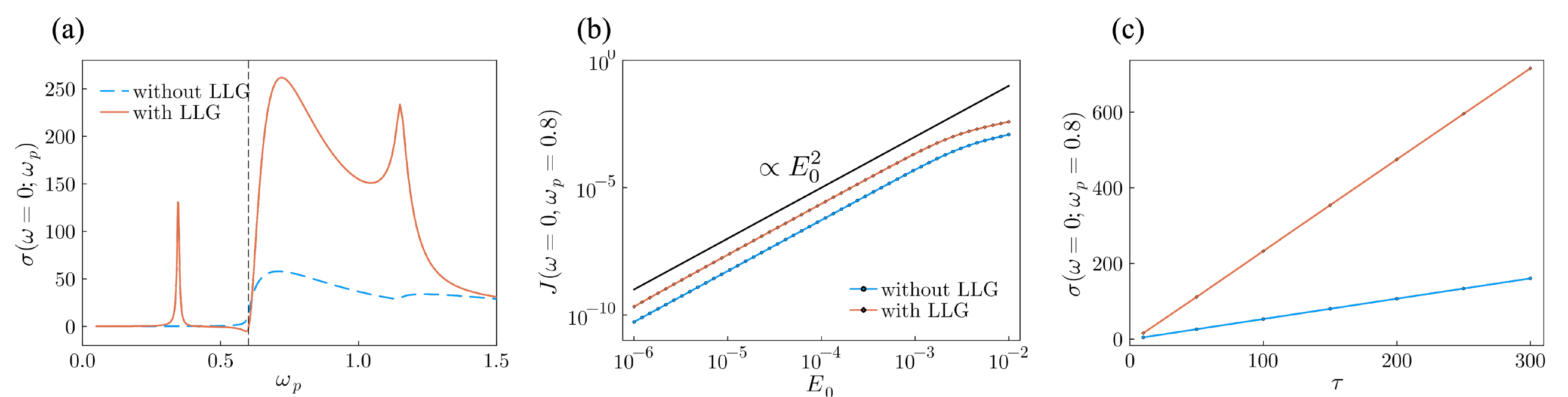}
        \caption{(a) Photocurrent spectra. The blue dashed and red solid lines indicate the IPA calculation and calculation incorporated the spin dynamics, respectively. (b,c) Dependence of photocurrent conductivity at the above bandgap frequency ($\omega_{p}=0.8$) on (b) electric-field strength $E_{0}$ and (c) relaxation time $\tau$. The blue and red lines indicated the IPA calculation and calculation incorporated the spin dynamics, respectively. The black solid line in (b) indicates the line which is proportional to $E_{0}^{2}$.}
        \label{PVE}
\end{figure*}
Now, we show the result of photocurrent spectra, revealing the effect of collective spin dynamics on the BPVE. The second-order BPVE can be described as \footnote{Here $\sigma^{z;zz}(0, \omega_{p}, -\omega_{p}) = \sigma^{z;zz}(0, -\omega_{p}, \omega_{p})^{\ast} = \sigma^{z;zz}(0, -\omega_{p}, \omega_{p})$ holds.}
\begin{align}
        \begin{split}
            J^{z}(\omega=0,\omega_{p}) = \sigma^{z;zz}(0;\omega_{p}, -\omega_{p})E^{z}(\omega_{p})E^{z}(-\omega_{p})\\+ \sigma^{z;zz}(0;-\omega_{p},\omega_{p})E^{z}(-\omega_{p})E^{z}(\omega_{p}).
        \end{split}
\end{align}
Here $J^{z}(\omega=0,\omega_{p}) $ is the DC component of output current along chain direction $z$ induced by the light field with frequency $\omega_{p}$. 
To calculate the photocurrent spectra, we apply the continuous light field $E^{z}(t) = E_{0}\sin\omega_{p} t$. 
Here we define the photocurrent conductivity $\sigma(0;\omega_{p})$ as 
\begin{align}
        \sigma(\omega=0;\omega_{p}) &= \dfrac{1}{2}\qty(\sigma^{z;zz}(0; \omega_{p}, -\omega_{p}) + \sigma^{z;zz}(0; -\omega_{p}, \omega_{p})).
\end{align}
In this definition, the photocurrent conductivity is calculated by the following formula \cite{Kaneko2021}
\begin{align}
\sigma(\omega=0; \omega_{p}) = \dfrac{2}{E_{0}^{2}NT_{p}}\int_{t_{\text{sat}}}^{t_{\text{sat}} + NT_{p}}J^{z}(t)dt.
\end{align}
Here $T_{p} = 2\pi/\omega_{p}$ is the period of the external light. We set $t_{\text{sat}}$ large enough so that the system reaches the time-periodic steady state and use $N > 10$ to get the average. 

In \figref{PVE}(a), we show the photocurrent spectra, with and without the contributions from spin dynamics. The blue dashed line indicates the IPA calculation, and the red solid line indicates the photocurrent spectra with the effect of spin dynamics.

Firstly, we observed the sharp resonant contribution from collective spin dynamics in the in-gap regime. This corresponds to the $\alpha$-mode resonance we observed in the linear response functions in \figref{linear_response_function}. Secondly, we confirmed the substantial enhancement above the band gap regime, which comes from off-resonant spin dynamics. Moreover, we observed the dip structure at the band edge frequency, which comes from the interference effect of the external field and internal spin dynamics, as shown below. Unlike the case of linear optical conductivity in \figref{linear_response_function}(b), there is no sum rule regarding photocurrent conductivity \cite{Watanabe2020_sum_rule}. Therefore, the photocurrent response in the in-gap regime is not subject to a trade-off relation with that in the above-gap.
In \figref{PVE} (b), (c), we show the field amplitude and the relaxation time $\tau = \gamma^{-1}$ dependence of the photocurrent at the above-band gap frequency $\omega_{p}=0.8$. 
Although the photocurrent response obtained in our method includes a higher-order nonlinear optical response to the light field in principle, we confirmed that the second-order response is dominant. Additionally, the relaxation time dependence of the photocurrent spectra scaling linearly to the relaxation time indicates that the injection current contribution is dominant, which agrees with symmetry analysis. It is noteworthy that our time-dependent calculation captures photocurrent not only from resonant contributions but also from off-resonant contributions of collective spin dynamics, which were not discussed in the earlier studies \cite{Morimoto2019_electromagnon, Morimoto2021} of BPVE from magnetic excitation. 

\subsection{Decomposition of Photocurrent}\label{decomposition_of_photocurrent}
\begin{figure}
        \centering
        \includegraphics[width = \linewidth]{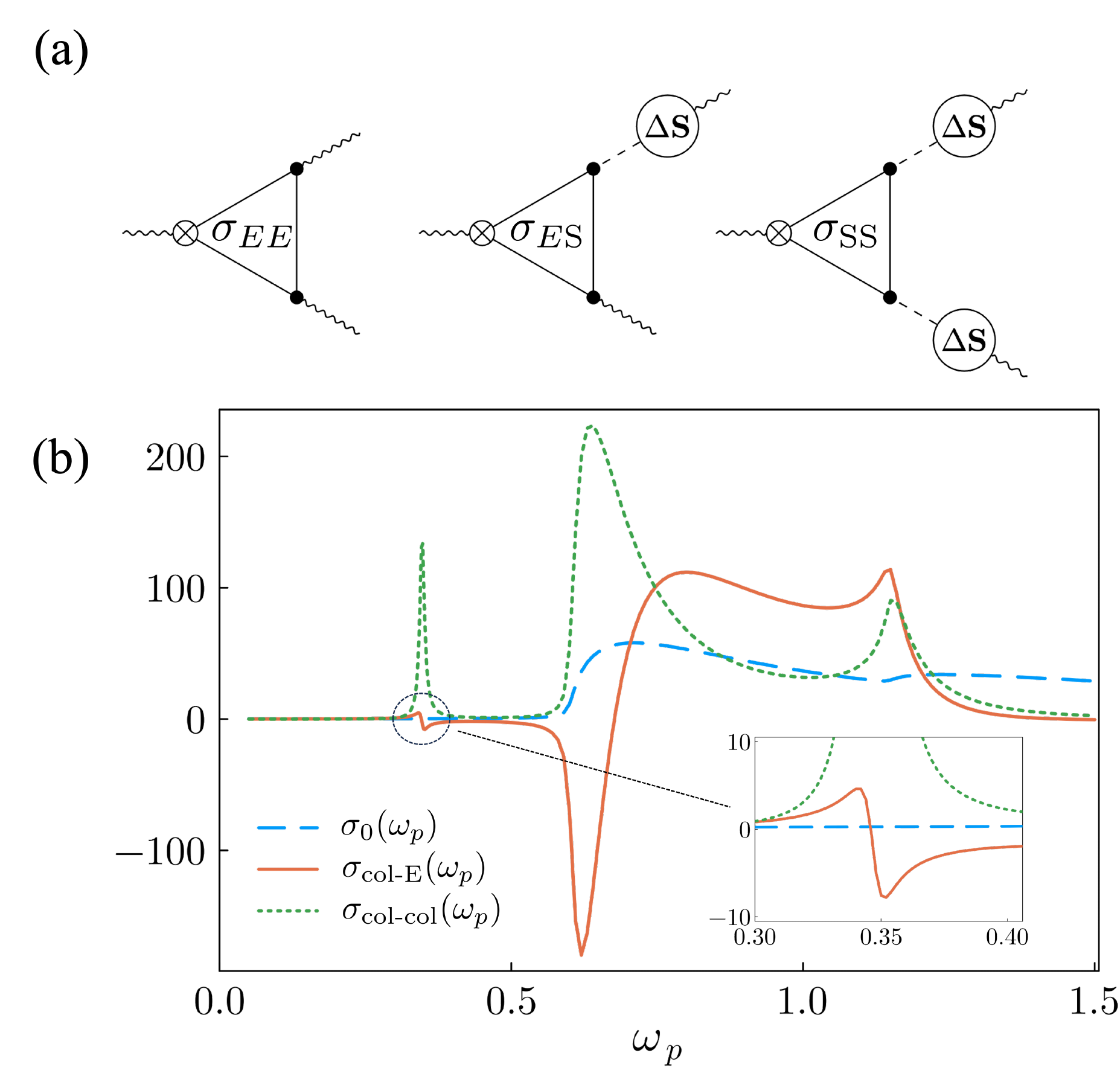}
        \caption{(a) Schematic picture of three different processes of the photocurrent in the presence of collective spin dynamics. The wavy and dashed lines indicate the light field and interaction $J$, respectively. $\otimes$ represents the output photocurrent.
        The solid triangles describe the photocurrent susceptibility evaluated with the IPA. $\Delta \vb{S}$ is the light induced spin dynamics.
        (b) Photocurrent spectra originating from the three different processes. The inset shows the magnified view of $\sigma_{\text{col-E}}$ component around the collective mode frequency $\omega_{p} \sim 0.35$.}
        \label{diag_decomp}
\end{figure}
\begin{figure*}
        \includegraphics[width = \linewidth]{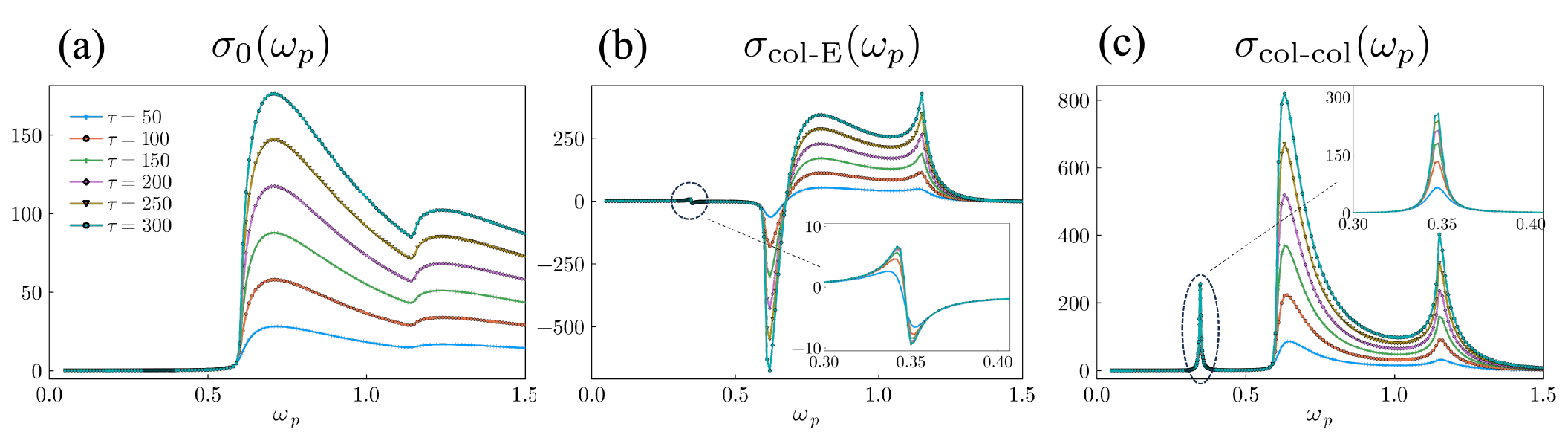}
        \caption{Relaxation time dependence of each photocurrent conductivity component. The insets in (b) and (c) show the magnified view of photocurrent spectra around the collective mode frequency $\omega_{p} \sim 0.35$. }
        \label{diagram_tau_dependence}
\end{figure*}  
To obtain further insight into the role of spin dynamics on the BPVE, we decompose the photocurrent contributions into three distinct processes. Our calculation is essentially equivalent to the random phase approximation \cite{Murakami2020, Kurebayashi2021}, wherein the light-induced internal spin dynamics can be considered as an additional external field applied to the conduction electrons (see \Eqref{Hund_Hamiltonian} and \Eqref{time_dependent_hamiltonian}). Consequently, as illustrated in \figref{diag_decomp}(a), three distinct processes contribute to the photocurrent response, described as
\begin{align}
        J^{z}(\omega=0;\omega_{p}) &= J^{z}_{0}(\omega_{p}) + J^{z}_{\text{col-E}}(\omega_{p}) +J^{z}_{\text{col-col}}(\omega_{p}),
\end{align}
where 
\begin{align}
        J_{0}^{z}(\omega_{p}) &= \sigma_{EE}^{z;zz}(0;-\omega_{p}, \omega_{p})E^{z}(-\omega_{p})E^{z}(\omega_{p}) \nonumber\\
        &+ \sigma_{EE}^{z;zz}(0;\omega_{p}, -\omega_{p})E^{z}(\omega_{p})E^{z}(-\omega_{p})\label{J_0},\\
        J_{\text{col-E}}^{z}(\omega_{p}) &= \sum_{\lambda}\sigma_{E\mathrm{S}}^{z;z\lambda}(0;-\omega_{p},\omega_{p})E^{z}(-\omega_{p})\Delta\mathrm{S}^{\lambda}(\omega_{p}) \nonumber\\ 
        &+\sum_{\lambda}\sigma_{E\mathrm{S}}^{z;z\lambda}(0;\omega_{p},-\omega_{p})E^{z}(\omega_{p})\Delta\mathrm{S}^{\lambda}(-\omega_{p}), \label{J_colE}\\
        J_{\text{col-col}}^{z}(\omega_{p}) &= \sum_{\nu\lambda}\sigma_{\mathrm{S}\mathrm{S}}^{z;\nu\lambda}(0;-\omega_{p},\omega_{p})\Delta \mathrm{S}^{\nu}(-\omega_{p})\Delta \mathrm{S}^{\lambda}(\omega_{p}) \nonumber \\
        &+ \sum_{\nu\lambda}\sigma_{\mathrm{S}\mathrm{S}}^{z;\nu\lambda}(0;\omega_{p},-\omega_{p})\Delta \mathrm{S}^{\nu}(\omega_{p})\Delta \mathrm{S}^{\lambda}(-\omega_{p}) \label{J_colcol}.
\end{align}
Here we use the spin dynamics $\Delta\vb{S}(\omega)$ defined in \Eqref{alpha_mode_dynamics}, and indices $\nu, \lambda$ represent the components of $\Delta\vb{S}(\omega)$. 
Firstly, $J_{0}$ depicts the photocurrent in the absense of collective spin dynamics, described by the photocurrent conductivity $\sigma_{EE}$. This contribution corresponds to the IPA. 
The second contribution is $J_{\text{col-E}}$, where the external light field and light-induced spin dynamics synergistically generate the photocurrent, and is characterized by $\sigma_{E\mathrm{S}}$. This can be understood as interference between the external light field and collective spin dynamics. The third contribution $J_{\text{col-col}}$ involves solely the light-induced internal spin fields, to generate the photocurrent, and is characterized by $\sigma_{\mathrm{SS}}$. Given that we have confirmed the linear coupling of spin dynamics to the light field, all these components yield a second-order response to the light field. 
We emphasize that $\sigma_{EE}, \sigma_{E\mathrm{S}}, \sigma_{\mathrm{SS}}$ are calculated within the IPA framework. The detailed expressions of them are given in \appref{pve_conductivity}.
In linear response, the total output current with multiple fields is simply the superposition of the individual outputs for each field. However, in nonlinear response, the total output is influenced by interference effects that result from the interaction of different fields.
Based on the photocurrent classification in \Eqref{J_0}, \Eqref{J_colE}, \Eqref{J_colcol}, we can define photocurrent conductivity in the presence of collective spin dynamics as  
\begin{align}
        J_{0}^{z}(\omega_{p}) &= 2\sigma_{0}(0;\omega_{p})E^{z}(\omega_{p})E^{z}(-\omega_{p}), \\
        J_{\text{col-E}}^{z}(\omega_{p}) &= 2\sigma_{\text{col-E}}(0;\omega_{p})E^{z}(\omega_{p})E^{z}(-\omega_{p}), \\
        J_{\text{col-col}}^{z}(\omega_{p}) &= 2\sigma_{\text{col-col}}(0;\omega_{p})E^{z}(\omega_{p})E^{z}(-\omega_{p}),
\end{align}
where 
\begin{align}
        \sigma_{0}(0;\omega_{p}) &= \dfrac{1}{2}\qty(\sigma_{EE}^{z;zz}(0; -\omega_{p}, \omega_{p}) + \sigma_{EE}^{z;zz}(0; \omega_{p}, -\omega_{p})), \\
        \begin{split}
        \sigma_{\text{col-E}}(0;\omega_{p})
        &=\dfrac{1}{2}\sum_{\lambda}\sigma_{E\mathrm{S}}^{z;z\lambda}(0;-\omega_{p},\omega_{p})\chi_{\mathrm{S}^{\lambda}E}(\omega_{p}) \\
        &+ \dfrac{1}{2}\sum_{\lambda}\sigma_{E\mathrm{S}}^{z;z\lambda}(0;\omega_{p},\omega_{p})\chi_{\mathrm{S}^{\lambda}E}(-\omega_{p}), 
        \end{split}\\
        \begin{split}
            \sigma_{\text{col-col}}(0;\omega_{p})
            &= \dfrac{1}{2}\sum_{\nu\lambda}\sigma_{\mathrm{SS}}^{z;\nu\lambda}(0;-\omega_{p},\omega_{p})\chi_{\mathrm{S}^{\nu}E}(-\omega_{p})\chi_{\mathrm{S}^{\lambda}E}(\omega_{p}) \\
        &+ \dfrac{1}{2}\sum_{\nu\lambda}\sigma_{\mathrm{SS}}^{z;\nu\lambda}(0;\omega_{p},-\omega_{p})\chi_{\mathrm{S}^{\nu}E}(\omega_{p})\chi_{\mathrm{S}^{\lambda}E}(-\omega_{p}).
        \end{split}
\end{align}
Here we use electromagnetic susceptibilities defined in \Eqref{electromagnetic_susceptibilities}.
Our time-dependent calculations naturally give the decomposition as follows. Firstly, $\sigma_{0}(\omega_{p})$ corresponds to the calculation without updating spin configurations. Secondly, $\sigma_{\text{col-col}}(\omega_{p})$ can be calculated by switching off the external light field $E^{z}(t)$ and updating the Hamiltonian with $\vb{S}(t)$ obtained in the calculations with the LLG equation. The $\sigma_{\text{col-E}}(\omega_{p})$ spectrum can be calculated by subtracting $\sigma_{0}(\omega_{p})$ and $\sigma_{\text{col-col}}(\omega_{p})$ from $\sigma(\omega_{p})$.

In \figref{diag_decomp}(b), we show the photocurrent spectra corresponding to each component. $\sigma_{0}(\omega_{p})$ is the same as the blue dashed line in \figref{PVE}(a). $\sigma_{\text{col-E}}(\omega_{p})$ component undergoes sign change near the bandgap frequency. This sign change may come from the interference effect between the external light field and field-induced spin dynamics, which might give a new way to control the photocurrent direction. 
In addition, we confirmed that the strong resonance peak mainly comes from $\sigma_{\text{col-col}}(\omega_{p})$ component. Since the resonant structure of $\sigma(\omega_{p})$ in the in-gap regime originates from the resonant structure in $\chi_{\vb{S}E}(\omega_{p})$, $\sigma_{\text{col-col}} \propto \qty(\chi_{\vb{S}E})^{2}$ gives a shaper resonance peak than $\sigma_{\text{col-E}} \propto \chi_{\vb{S}E}$ does. 

\begin{table}[]
        \caption{Photocurrent generation induced by the light field and spin dynamics. For instance, light field $E$ and uniform spin dynamics along $y$ direction $\mathrm{M}^{y}$ induces $\sigma_{\text{shift}}$ and $\sigma_{\text{Enj}}$. $\sigma_{\text{shift}}$ is induced by in-phase component of $E$ and $\mathrm{M}^{y}$, while the $\sigma_{\text{Enj}}$ is induced by out-of-phase component of $E$ and $\mathrm{M}^{y}$. The detailed expression for $\sigma_{\text{shift}}, \sigma_{\text{gyro}}, \sigma_{\text{Enj}}, \sigma_{\text{Mnj}}$ are given in the \appref{pve_conductivity}}
        \begin{ruledtabular}
        \begin{tabular}{ccccc}
& $E$ & $\mathrm{L}^{x}$ & $\mathrm{M}^{y}$ & $\mathrm{L}^{z}$ \\ \hline
$E$ & $\sigma_{\text{Mnj}}$  & $\sigma_{\text{Mnj}}$, $\sigma_{\text{gyro}}$ & $\sigma_{\text{shift}}$, $\sigma_{\text{Enj}}$ & $\sigma_{\text{shift}}$, $\sigma_{\text{Enj}}$ \\
$\mathrm{L}^{x}$ & $\sigma_{\text{Mnj}}, \sigma_{\text{gyro}}$ & $\sigma_{\text{Mnj}}$ & $\sigma_{\text{shift}}, \sigma_{\text{Enj}}$ & $\sigma_{\text{shift}}, \sigma_{\text{Enj}}$ \\
$\mathrm{M}^{y}$ & $\sigma_{\text{shift}}, \sigma_{\text{Enj}}$ & $\sigma_{\text{shift}}, \sigma_{\text{Enj}}$ & $\sigma_{\text{Mnj}}$ &  $\sigma_{\text{Mnj}}$         \\
$\mathrm{L}^{z}$ & $\sigma_{\text{shift}}, \sigma_{\text{Enj}}$ & $\sigma_{\text{shift}}, \sigma_{\text{Enj}}$ & $\sigma_{\text{Mnj}}$  & $\sigma_{\text{Mnj}}$  
\end{tabular}
\end{ruledtabular}   
\label{photocurrent_classification}
\end{table} 

It is worth noting that symmetry relations and the phase degrees of freedom of driving fields constrain the types of photocurrent generated. As a general case, let us focus on noninteracting electrons and consider the photocurrent induced by two external fields $X(\omega_{p})$ and $Y(\omega_{p})$ within the IPA framework. This process can be written by using the
photocurrent conductivity $\sigma_{XY}(0; \omega_{p}, -\omega_{p})$ as
\begin{align}
    \begin{split}
        J_{XY}(\omega_{p}) &= \sigma_{XY}(0;-\omega_{p}, \omega_{p})X(-\omega_{p})Y(\omega_{p})\\
     &+  \sigma_{XY}(0; \omega_{p}, -\omega_{p})X(\omega_{p})Y(-\omega_{p}).
    \end{split}
\end{align}
As we show in \appref{pve_conductivity}, $\sigma_{XY}$ can be classified into the following four components,
\begin{align}
    \sigma_{XY} = \sigma_{XY, \text{shift}} + \sigma_{XY, \text{gyro}} + \sigma_{XY, \text{Mnj}} + \sigma_{XY, \text{Enj}}. \label{photocurrent_XY}
\end{align}
Here, $\sigma_{XY, \text{shift}}$ and $\sigma_{XY, \text{gyro}}$ correspond to shift current and gyration current, while $\sigma_{XY, \text{Mnj}}$ and $\sigma_{XY, \text{Enj}}$ correspond to magnetic injection current and electric injection current, as discussed in the context of normal photocurrent \cite{Watanabe2021, Ahn2020}. These contributions are different in terms of relaxation time dependence. $\sigma_{XY, \text{shift}}$ and $\sigma_{XY, \text{gyro}}$ are independent of relaxation time, whereas $\sigma_{XY, \text{Mnj}}$ and $\sigma_{XY, \text{Enj}}$ are proportional to relaxation time. 
Additionally, $\sigma_{XY, \text{shift}}, \sigma_{XY, \text{Mnj}}$ ($\sigma_{XY, \text{gyro}}, \sigma_{XY, \text{Enj}}$) is the real (pure-imaginary) quantities, which describes the counterpart of the photocurrent induced by linearly (circularly) polarized light. Therefore, $\sigma_{XY, \text{shift}}, \sigma_{XY, \text{Mnj}}$ are finite only when two fields $X(\omega_{p}), Y(\omega_{p})$ are in-phase to each other, while $\sigma_{XY, \text{gyro}}, \sigma_{XY, \text{Enj}}$ are finite only when $X(\omega_{p}), Y(\omega_{p})$ are out-of-phase to each other. Besides this restriction from phase degrees of freedom, the symmetry of the two fields also restricts the type of photocurrent. 

 In our case, the field $X(\omega_{p}), Y(\omega_{p})$ corresponds to the light field $E(\omega_{p})$ and/or the spin dynamics $\mathrm{S}^{\nu}(\omega_{p})$.  For instance, we confirmed that the spin dynamics $\mathrm{L}^{x}(\omega_{p})$ and $\mathrm{M}^{y}(\omega)$ has both in-phase and out-of-phase components, and their sign under $\theta 2_{x}$ are different. This fact restricts the photocurrent generation from these fields to shift current and electric injection current similar to the photocurrent response to time-reversal-symmetric systems \cite{Watanabe2021}. 
A similar argument can be applied to the case of the photocurrent originating from the interference of light field and collective spin dynamics described by $\sigma_{E\mathrm{S}}$. 
We summarize the photocurrent classification in \tabref{photocurrent_classification}.

From this analysis, we reveal that unlike $\sigma_{0}$,  $\sigma_{\mathrm{SS}}$ and $\sigma_{E\mathrm{S}}$ can include the component of shift current-like contribution. Moreover, although we consider the photocurrent induced by the linearly polarized light, $\sigma_{\mathrm{SS}}$ and $\sigma_{E\mathrm{S}}$ has $\sigma_{\text{Enj}}, \sigma_{\text{gyro}}$ component which is the counterpart of circularly polarized induced photocurrent.
Practically, we can observe the in-gap resonant structure in $\sigma_{\text{col-E}}(\omega_{p})$ is mainly coming from the injection current-like contribution, since $\operatorname{Re}\chi_{\mathrm{S}^{\nu}E}$ ($\operatorname{Im}\chi_{\mathrm{S}^{\nu}E}$) represents the in-phse (out-of-phase) component of spin dynamics to the light field  (\appref{sym_analysis_photocurrent} for a more detailed discussion). Here, we reveal that the fictitious fields arising from the spin collective dynamics result in various types of photocurrent, which is not allowed in the IPA. 

In \figref{diagram_tau_dependence}, we show the relaxation time $\tau$ dependence of photocurrent spectra. Photocurrent spectra increase linearly with relaxation time, indicating the injection current-like contribution is dominant. Although we have revealed the presence of shift current-like contribution in $\sigma_{\text{col-E}}$ and $\sigma_{\text{col-col}}$, the primary contributions to $\sigma_{\text{col-E}}$ and $\sigma_{\text{col-col}}$ are injection current-like one, namely $\sigma_{\text{Enj}}, \sigma_{\text{Mnj}}$.

\subsection{Tuning of photocurrent by external magnetic field}\label{Tuning_of_photocurrent}
\begin{figure}
        \centering
        \includegraphics[width=\linewidth]{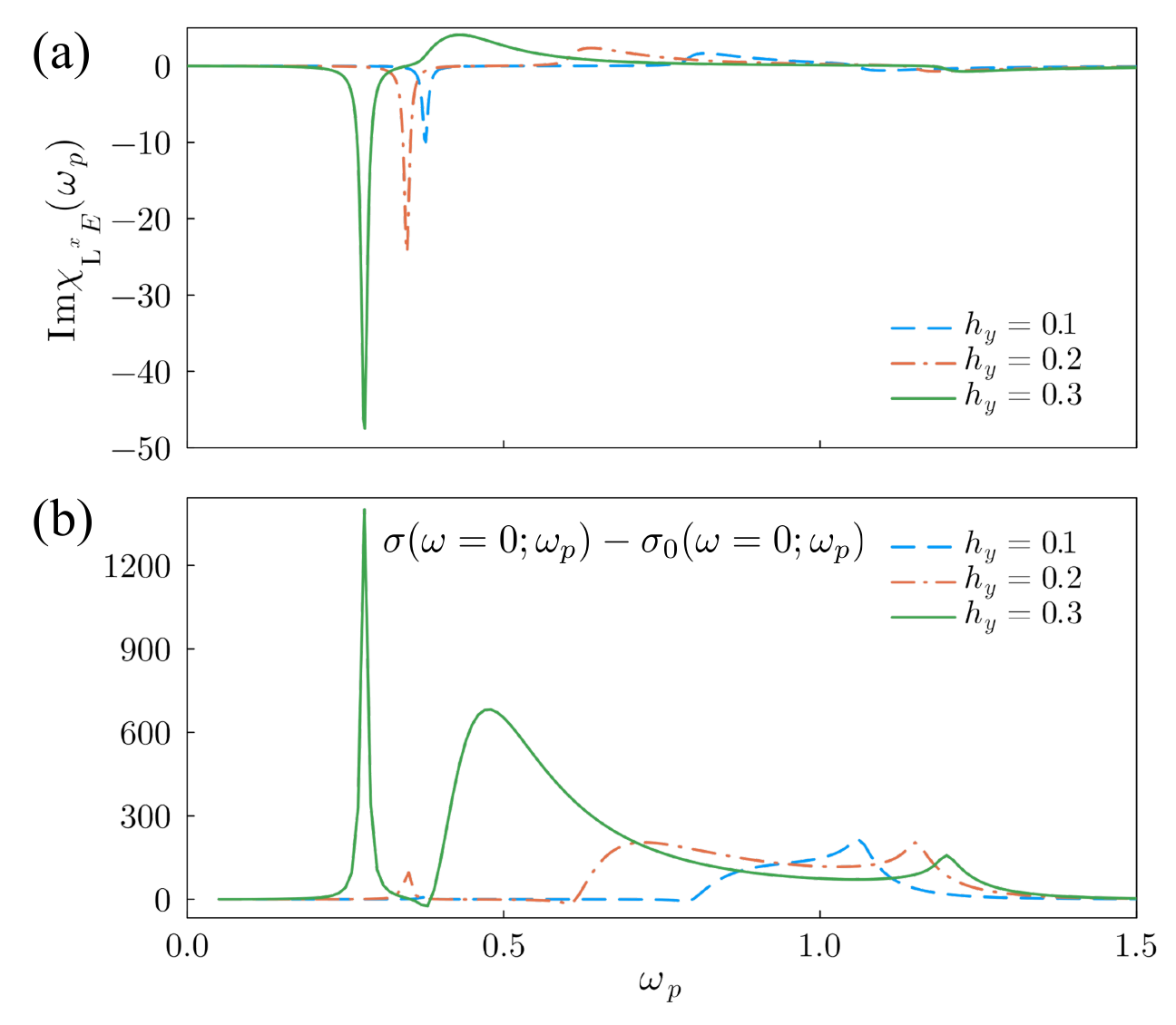}
        \caption{(a) Magnetic field dependence of linear electromagnetic susceptibility $\operatorname{Im}\chi_{\mathrm{L}^{x}E}(\omega)$ to the external light field.  (b) Photocurrent spectra $\sigma(\omega=0;\omega_{p}) - \sigma_{0}(\omega=0;\omega_{p})$ with modulating the canted angle of antiferromagnetic moments by changing the magnetic field along the $y$ direction}
        \label{pve_hy}
\end{figure}
Here, we demonstrate the tunability of photocurrent by applying the external magnetic field perpendicular to the chain direction, changing the canted angle of antiferromagnetic moments. In \figref{pve_hy}, we show $\vb*{\chi}_{\vb{S}E}(\omega_{p})$ and the photocurrent spectra originating from spin dynamics by changing the amplitude of the external magnetic field. 

In \figref{pve_hy} (a), we show the linear electromagnetic susceptibility of $\operatorname{Im}\chi_{\mathrm{L}^{x}E}$. As we increase the magnetic field along the $y$ axis, the amplitude of $\operatorname{Im}\chi_{\mathrm{L}^{x}E}$ gradually increases, showing the strong modulation of feedback between charge and spin degrees of freedom by the external field.
This modulation arises from the fact that the alteration of the canted angle induces changes in the optical gap of the system and the energy scale of the collective spin mode. Consequently, this modulation allows us to tune the strength of the feedback from spin dynamics.

Consistent with the modulation of the electromagnetic susceptibility, the photocurrent conductivity spectra also experience large modulation due to the external field, as shown in \figref{pve_hy} (b).
As explained in the previous subsection, $\vb*{\chi}_{\vb{S}E}(\omega_{p})$ plays a critical role in photocurrent generation in the presence of collective spin dynamics. Therefore, larger $\vb*{\chi}_{\vb{S}E}(\omega_{p})$ results in a more significant modulation in photocurrent generation.

As these results demonstrate the importance of energy scale matching between collective modes and electronic excitations, we can expect substantial modulation of photocurrent from magnetic excitation in the narrow gap system with the higher collective mode frequencies. 

\section{Summary and discussion}
We investigated the effect of antiferromagnetic excitations on linear and nonlinear optical responses by a real-time calculation method. Firstly, we observed the optical response arising from an electrically active antiferromagnetic resonance mode, which aligned with our symmetry analysis. 
We further found that collective spin dynamics, originating from both resonant and off-resonant contributions, can significantly enhance the photocurrent response. In addition, we delved into a comprehensive analysis by classifying it into different processes. The component includes the unique photocurrent arising from the interference between the light field and the internal spin field, which is inherent to the nonlinear response. 
Moreover, based on the symmetry analysis, we revealed that the emergence of the spin dynamics allows the various types of photocurrent, which are totally absent in the IPA. Additionally, we showcased the tunability of the photocurrent in our system by varying the canted angles and emphasized the significance of matching in the energy scales of collective modes and electronic excitations.

Now, we estimate the strength of modulation by spin dynamics. We asuume the energy scale of hopping $t_{h} = \SI{0.1}{eV}$ and lattice constant $a = \SI{0.5}{nm}$. 
Additionally, to compare the obtained photocurrent conductivity in a one-dimensional system to that of a three-dimensional system, we assume the three-dimensional stack of one-dimensional chains with a lattice spacing of $\SI{1.0}{nm}$ \cite{Cook2017}, and realistic spin-orbit coupling $\lambda = 0.1$ \cite{Smidman2017} (see \appref{photocurrent_estimation}). In this condition, the modulation of the photocurrent conductivity by spin dynamics $\sigma_{\text{col-E}} + \sigma_{\text{col-col}}$ reaches the order of $\sim\SI{1.0}{mA/V^2}$, which can be comparable to that of Weyl semimetals \cite{Osterhoudt2019}, demonstrating the significance of the effect of spin dynamics on photocurrent response.

While our calculations are limited to the simple two-sublattice model, we can extend these to more complex systems, like skyrmions \cite{Nagaosa2013, Tokura2021, Zhang2016}, and hedgehogs \cite{Okumura_hedgehog}. The essential ingredient is the optically active collective spin dynamics, whose energy scale is close to electronic degrees of freedom. Therefore, searching for a real material should be in the realm of inversion broken narrow gap semiconductors with optically active magnons, such as $\ce{MnBi2Te4}$ \cite{Li2020, Fei2020}. 

\section*{acknowledgement}
This work is supported by Grant-in-Aid for Scientific Research from JSPS, KAKENHI Grant No.~JP23K13058 (H.W.), No.~JP21H05017 (Y.M.), No.~21H04990 (R.A.), No.~JP21H04437 (R.A.), No.~JP20K14412 (Y.M.), No.~19H05825 (R.A.), JST-PRESTO No.~JPMJPR20L7 (T.N.), and JST-CREST No.~JPMJCR1901 (Y.M.), No.~JPMJCR18T3 (R.A.).

\appendix

\section{$J, \lambda$-dependence of the photocurrent spectra}\label{photocurrent_estimation}
\begin{figure}
    \centering
    \includegraphics[width=\linewidth]{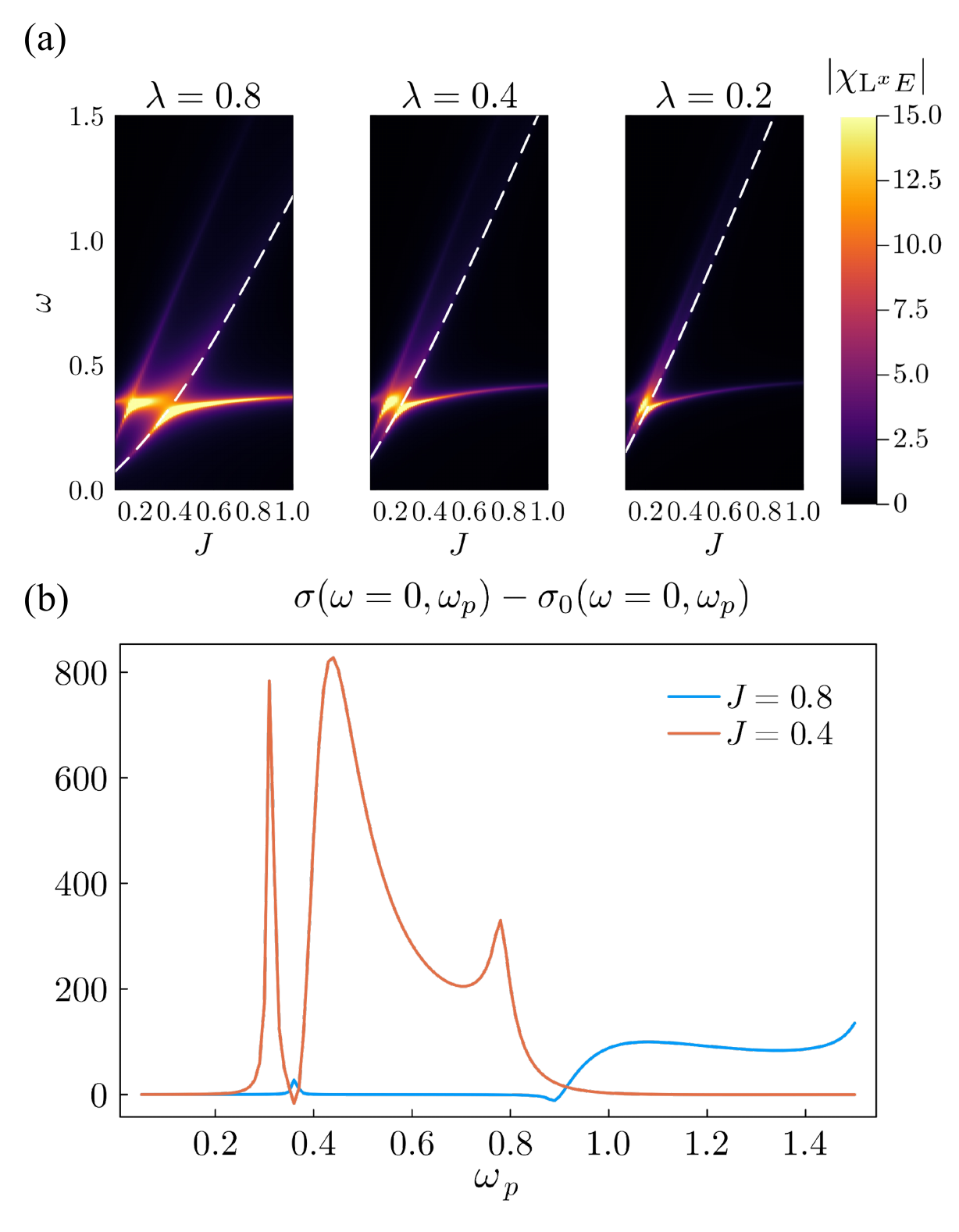}
    \caption{(a) $J, \lambda$-dependence of amplitude of electromagnetic susceptibility. The white dashed lines indicate the band. The parameters $t_{h} = 1, K_{z} = 0.2, h_{z} = 0.2, \gamma = 0.01$ were used. (b) Photocurrent modulation by spin dynamics $\sigma(\omega=0;\omega_{p}) - \sigma_{0}(\omega=0;\omega_{p})$
    with the parameters $t_h=1,\lambda=0.8, K_z=0.2,h_y=0.2$.}
    \label{pve_J_dep}
\end{figure}
\begin{figure}
    \centering
    \includegraphics[width=\linewidth]{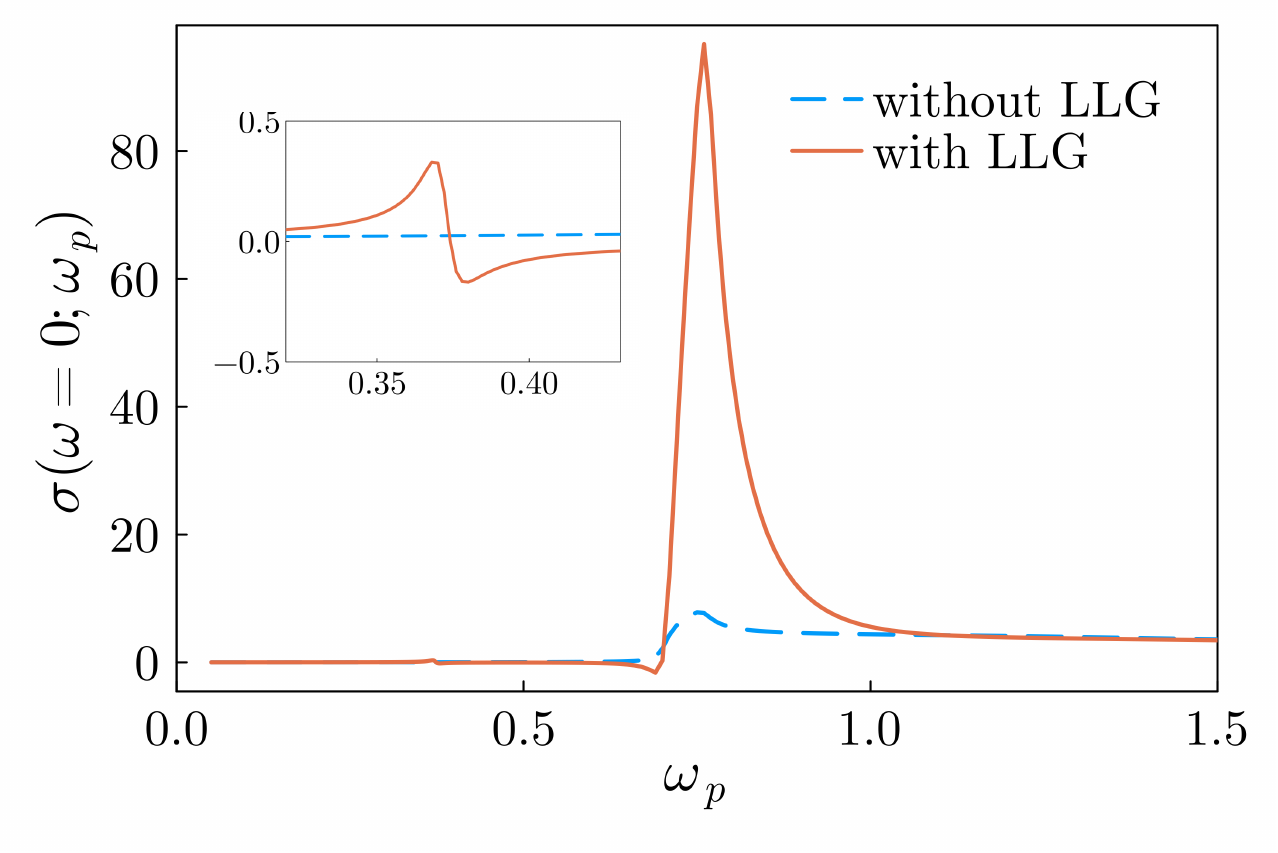}
     \caption{Photocurrent spectra with smaller spin-orbit coupling. The parameters $t_{h} = 1, \lambda = 0.1, J = 0.4, K_{z} = 0.2, h_{z} = 0.2, \gamma =
0.01$ were used. The blue dashed and red solid lines indicate the IPA calculation and calculation incorporated the spin dynamics, respectively.}
    \label{pve_appendix}
\end{figure}
Here, we supplementally discuss the effect of collective spin dynamics on BPVE by changing the strength of spin-charge coupling $J$ and spin-orbit coupling $\lambda$. 
As we discussed in the main text, electromagnetic susceptibility $\chi_{\vb{S}E}(\omega)$ determines the spin dynamics correction to the photocurrent response. In \figref{pve_J_dep}, we show $J$ and $\lambda$-dependence of $\abs{\chi_{\mathrm{L}^{x}E}(\omega)}$. White dashed lines indicate the band gap energy of the system. Firstly, by changing $J$, the resonance peak of $\abs{\chi_{\mathrm{L}^{x}E}(\omega)}$ is more prominent when the energy scale of the spin resonance and the electronic band gap are proximate to each other. A critical insight here is that altering $J$ affects both the electronic structure and the frequency of collective modes, leading to a non-trivial relationship where larger $J$ does not necessarily yield larger feedback from spin dynamics. Therefore, as discussed in the main text, the energy scale proximity of the electronic and spin systems is important to obtain a large response from the spin dynamics. This observation aligns with the result of photocurrent modulation by spin dynamics in \figref{pve_J_dep} (b). Here, moderate exchange coupling facilitates energy-scale proximity between the electronic band gap and collective mode frequency, thereby enhancing photocurrent generation from spin dynamics.
Additionally, comparing the $\abs{\chi_{\mathrm{L}^{x}E}}(\omega)$ with different spin-orbit coupling in \figref{pve_J_dep} (a), we can see the tendency that the value of spin-orbit coupling $\lambda$ determines the amplitude of the light field-induced spin dynamics. In the case of large spin-orbit coupling, we observe a large amplitude of light-induced spin dynamics. Note that with smaller $J$, we cannot distinguish the in-gap resonant excitation from the above-gap off-resonant excitation since the resonant frequency of local spin dynamics overlaps the continuum excitation of electronic systems.

Furthermore, we estimate the photocurrent generation with the smaller value of spin-orbit coupling to make a realistic estimation.
In \figref{pve_appendix}, we show the photocurrent spectra with and without the effect of spin dynamics. Here the parameters $t_{h} = 1, \lambda = 0.1, J = 0.4, K_{z} = 0.2, h_{z} = 0.2, \gamma = 0.01$ are used. Contrary to the photocurrent spectra in \figref{PVE} (a), the resonant structure below the bandgap regime is suppressed due to the smaller magnetoelectric coupling. However, even with the small spin-orbit coupling, we can see significant modification from off-resonant spin dynamics above the band gap spectra. Using these parameters and assuming the relaxation time $\tau \sim \SI{60}{fs}$, we can estimate that the enhanced photocurrent response can reach the order of $\sim\SI{1.0}{mA/V^2}$, which is comparable to Weyl semimetal \cite{Osterhoudt2019}. 

\onecolumngrid

\section{Nonlinear optical response functions based on density matrix formalism}\label{pve_conductivity}
This section is dedicated to showing the formalism of perturbative expansion of the von Neumann equation \Eqref{vonNeumann} in the presence of both the light field and light-induced spin dynamics. 
The Hamiltonian for electrons under an external light field and effective magnetic fields from the spin dynamics is expressed as 
\begin{align}
        \mathcal{H}(t) = \mathcal{H}_{0} + \Delta\mathcal{H}(t).
\end{align}
$\Delta\mathcal{H}(t)$ is the perturbative Hamiltonian and consists of an external light field and light-induced spin dynamics defined as
\begin{align}
        \Delta\mathcal{H}(t) &= \Delta\mathcal{H}_{E}(t) + \Delta\mathcal{H}_{\text{spin}}(t), \\
        \Delta\mathcal{H}_{E} &= -r^{\mu}E^{\mu}, \label{lengthgauge}\\
        \Delta\mathcal{H}_{\text{spin}}(t) &= -JA^{\mu}\Delta\mathrm{S}^{\mu}(t), \\
        \Delta\vb{S}(t) &= \qty(\Delta\mathrm{L}^{x}(t), \Delta\mathrm{M}^{y}(t), \Delta\mathrm{L}^{z}(t))^{T}, \\
        \vb*{A} &=  \qty(\sigma_{x}\tau_{z}, \sigma_{y}\tau_{0}, \sigma_{z}\tau_{z})^{T}.
\end{align}
Here, $\sigma^{\mu}, \tau^{\mu}$ are the Pauli matrices that describe the spin and sublattice degrees of freedom, respectively. 
$\mathcal{H}_{0}$ is the nonperturbative Hamiltonian given by
\begin{align}
        \mathcal{H}_{0} = \int \dfrac{d\vb*{k}}{(2\pi)^{d}}\sum_{a}\epsilon_{\vb*{k}a}c_{\vb*{k}a}^{\dagger}c_{\vb*{k}a}.
\end{align}
Here $c_{\vb*{k}a}^{\dagger} (c_{\vb*{k}a})$ is the creation (annihilation) operators of the Bloch state $\ket{\psi_{\vb*{k}a}} = \exp(i\vb*{k}\cdot \vb{r})\ket{u_{\vb*{k}a}}$. $\ket{u_{\vb*{k}a}}$ is the cell-periodic part of the Bloch state and satisfies the following equation.
\begin{align}
        H_{0}(\vb*{k})\ket{u_{\vb*{k}a}} = \epsilon_{\vb*{k}a}\ket{u_{\vb*{k}a}}.
\end{align}
According to \cite{ADAMS1959286}, in the infinite volume limit the position operator in \Eqref{lengthgauge} is written in the Bloch representation as
\begin{align}
        \qty[\vb*{r_{k}}]_{ab} = i\nabla_{\vb{k}}\delta_{ab} + \vb*{\xi}_{ab}. \label{position_operator}
\end{align}
Here $\vb*{\xi}_{ab}$ is the Berry connection defined as $\vb*{\xi}_{ab} = i\mel{u_{\vb*{k}a}}{\nabla_{\vb*{k}}}{u_{\vb*{k}b}}$. In the following calculation, $k$ dependence is omitted, otherwise explicitly mentioned.

As we explained in the main text, the von Neumann equation of SPDM $\rho_{ab} = \langle c_{\vb*{k}b}^{\dagger}c_{\vb*{k}a}\rangle $ is given by, 
\begin{align}
        i\pdv{\rho_{ab}(t)}{t} &= \qty[\mathcal{H}, \rho]_{ab} =\qty[\mathcal{H}_{0}, \rho]_{ab} + \qty[\Delta\mathcal{H}(t), \rho]_{ab}.
\end{align}
We define the Fourier transformation as
\begin{align}
        f(t) = \int \dfrac{d\omega}{2\pi}f(\omega)e^{-i(\omega + i\eta)t},
\end{align}
and we get the frequency-space representation of the von Neumann equation,
\begin{align}
       (\omega + i\eta - \epsilon_{ab})\rho_{ab}(\omega) =  \int \dfrac{d\omega_{1}}{2\pi}\qty[\Delta \mathcal{H}(\omega_{1}), \rho(\omega - \omega_{1})]_{ab}.
\end{align}
Here, we introduced the infinitesimal parameter $\eta$ to describe the adiabatic application of the external field.
We can perturbatively expand this equation by introducing the $\rho^{(n)}$, which is the $n$-th order with respect to the external field, 
\begin{align}
        (\omega + i\eta - \epsilon_{ab})\rho_{ab}^{(n+1)}(\omega) = \int\dfrac{d\omega_{1}}{2\pi}\qty[\Delta \mathcal{H}(\omega_{1}), \rho^{(n)}(\omega - \omega_{1})]_{ab}  \label{vonNeumann_omega}.
\end{align}
We introduce matrix $d(\omega)$ as 
\begin{align}
        d_{ab}(\omega) = \dfrac{1}{\omega + i\eta - \epsilon_{ab}}
        = \mathcal{P}\dfrac{1}{\omega-\epsilon_{ab}} - i\pi\delta(\omega-\epsilon_{ab}), \label{Dirac_formula}
\end{align}
and by using the Hadamard product $(A \odot B)_{ab} = A_{ab}B_{ab}$ \cite{Ventura2017}, the 
von Neumann equation \Eqref{vonNeumann_omega} can be recast as 
\begin{align}
        \rho_{ab}^{(n+1)}(\omega) = \int \dfrac{d\omega_{1}}{2\pi}\qty(d(\omega) \odot [\Delta \mathcal{H}(\omega_{1}), \rho^{(n)}(\omega - \omega_{1})])_{ab}.
\end{align}
In the following calculation, we solve this equation with the boundary condition of $\rho_{ab}^{(0)}(\omega) = 2\pi\delta(\omega)f_{ab}\delta_{ab}$. Since we focus on the second-order response to the external field, we solve the equation
\begin{align}
        \begin{split}
            \rho_{ab}^{(2)}(\omega) &= \int \dfrac{d\omega_{1}}{2\pi}\qty(d(\omega) \odot [\Delta \mathcal{H}(\omega_{1}), \rho^{(1)}(\omega - \omega_{1})])_{ab} \\
        &= \int \dfrac{d\omega_{1} d\omega_{2}}{(2\pi)^{2}}\qty(d(\omega) \odot [\Delta \mathcal{H}(\omega_{1}), d(\omega - \omega_{1})\odot [\Delta \mathcal{H}(\omega_{2}), \rho^{(0)}(\omega-\omega_{1}-\omega_{2})]])_{ab}.
        \end{split}
\end{align}
Considering both the light field and the spin dynamics, $\rho_{ab}^{(2)}(\omega)$ can be divided into three components as, 
\begin{align}
        \rho_{ab}^{(2)}(\omega) = \rho_{EE, ab}(\omega) + \rho_{E\mathrm{S}, ab}(\omega) + \rho_{\mathrm{S}\mathrm{S}, ab}(\omega),
\end{align}
where
\begin{align}
        \rho_{EE,ab}(\omega) &= \dfrac{1}{2}\int \dfrac{d\omega_{1} d\omega_{2}}{(2\pi)^{2}}E^{\nu}(\omega_{1})E^{\lambda}(\omega_{2})\qty(d(\omega) \odot [r^{\nu}, d(\omega - \omega_{1})\odot [r^{\lambda}, \rho^{(0)}(\omega-\omega_{1}-\omega_{2})]])_{ab} + [(\nu, \omega_{1}) \leftrightarrow (\lambda, \omega_{2})], \\
        \begin{split}
            \rho_{E\mathrm{S}, ab}(\omega) &= \dfrac{J}{2}\int \dfrac{d\omega_{1} d\omega_{2}}{(2\pi)^{2}}E^{\nu}(\omega_{1})\mathrm{S}^{\lambda}(\omega_{2})\qty(d(\omega) \odot [r^{\nu}, d(\omega - \omega_{1})\odot [A^{\lambda}, \rho^{(0)}(\omega-\omega_{1}-\omega_{2})]])_{ab} \\
        &+ \dfrac{J}{2}\int \dfrac{d\omega_{1} d\omega_{2}}{(2\pi)^{2}}\mathrm{S}^{\lambda}(\omega_{1})E^{\nu}(\omega_{2})\qty(d(\omega) \odot [A^{\lambda}, d(\omega - \omega_{1})\odot [r^{\nu}, \rho^{(0)}(\omega-\omega_{1}-\omega_{2})]])_{ab},
        \end{split}\\
        \rho_{\mathrm{S}\mathrm{S}, ab}(\omega) &= \dfrac{J^{2}}{2}\int \dfrac{d\omega_{1} d\omega_{2}}{(2\pi)^{2}}\mathrm{S}^{\nu}(\omega_{1})\mathrm{S}^{\lambda}(\omega_{2})\qty(d(\omega) \odot [A^{\nu}, d(\omega - \omega_{1})\odot [A^{\lambda}, \rho^{(0)}(\omega-\omega_{1}-\omega_{2})]])_{ab} + [(\nu, \omega_{1}) \leftrightarrow (\lambda, \omega_{2})].
\end{align}
Here, we symmetrize the expression of $\rho$ since the perturbative calculation should be invariant under the arbitrary exchange of external field. Based on this decomposition, we formulate the photocurrent formula in the presence of spin dynamics.
\subsection{Light field induced photocurrent}
Here we derive the light field-induced photocurrent based on the $\rho_{EE}$. It is convenient to decompose the position operator in \Eqref{position_operator} into inter- ($\vb*{r}_{e}$) and intra-band ($\vb*{r}_{i}$) component as
\begin{align}
        \begin{split}
            \qty(\vb*{r}_{i})_{ab} &= \delta_{ab}\qty(i\nabla_{\vb{k}} + \xi_{aa}), \\
            \qty(\vb*{r}_{e})_{ab} &= \qty(1-\delta_{ab})\xi_{ab}.  
        \end{split}    
\end{align}
Based on this decomposition, $\rho_{EE}(\omega)$ can be classified into the following four component, 
\begin{align}
        \rho_{EE,ab}(\omega) = \rho_{EE, ab}^{(ii)} + \rho_{EE,ab}^{(ie)} + \rho_{EE, ab}^{(ei)} + \rho_{EE,ab}^{(ee)}.
\end{align}
Each term is explicitly written as,
\begin{align}
        \begin{split}
            \rho_{EE, ab}^{(ii)} &= \dfrac{1}{2}\int \dfrac{d\omega_{1} d\omega_{2}}{(2\pi)^{2}}E^{\nu}(\omega_{1})E^{\lambda}(\omega_{2})\qty(d(\omega) \odot [r_{i}^{\nu}, d(\omega - \omega_{1})\odot [r_{i}^{\lambda}, \rho^{(0)}(\omega-\omega_{1}-\omega_{2})]])_{ab} + [(\nu, \omega_{1}) \leftrightarrow (\lambda, \omega_{2})] \\
            &= -\int \dfrac{d\omega_{1}d\omega_{2}}{(2\pi)^{2}}E^{\nu}(\omega_{1})E^{\lambda}(\omega_{2})d_{ab}(\omega)d_{ab}(\omega-\omega_{1})\partial_{\nu}\partial_{\lambda}f(\epsilon_{\vb*{k}a})\delta_{ab}2\pi\delta(\omega-\omega_{1}-\omega_{2}) + [(\nu, \omega_{1}) \leftrightarrow (\lambda, \omega_{2})], 
        \end{split} \\
        \begin{split}
            \rho_{EE, ab}^{(ie)} &= \dfrac{1}{2}\int \dfrac{d\omega_{1} d\omega_{2}}{(2\pi)^{2}}E^{\nu}(\omega_{1})E^{\lambda}(\omega_{2})\qty(d(\omega) \odot [r_{i}^{\nu}, d(\omega - \omega_{1})\odot [r_{e}^{\lambda}, \rho^{(0)}(\omega-\omega_{1}-\omega_{2})]])_{ab} + [(\nu, \omega_{1}) \leftrightarrow (\lambda, \omega_{2})]  \\
            &= -\dfrac{i}{2}\int\dfrac{d\omega_{1}\omega_{1}}{(2\pi)^{2}}E^{\nu}(\omega_{1})E^{\lambda}(\omega_{2})d_{ab}(\omega)\qty[\partial_{\nu}\qty(d_{ab}(\omega-\omega_{1})\xi_{ab}^{\lambda}f_{ab})-i\qty(\xi_{aa}^{\nu} - \xi_{bb}^{\nu})d_{ab}(\omega-\omega_{1})\xi_{ab}^{\lambda}f_{ab}] + [(\nu, \omega_{1}) \leftrightarrow (\lambda, \omega_{2})]
        \end{split},\\
        \begin{split}
            \rho_{EE, ab}^{(ei)} &= \dfrac{1}{2}\int \dfrac{d\omega_{1} d\omega_{2}}{(2\pi)^{2}}E^{\nu}(\omega_{1})E^{\lambda}(\omega_{2})\qty(d(\omega) \odot [r_{e}^{\nu}, d(\omega - \omega_{1})\odot [r_{i}^{\lambda}, \rho^{(0)}(\omega-\omega_{1}-\omega_{2})]])_{ab} + [(\nu, \omega_{1}) \leftrightarrow (\lambda, \omega_{2})]  \\
            &=-\dfrac{i}{2}\int \dfrac{d\omega_{1}d\omega_{2}}{(2\pi)^{2}}E^{\nu}(\omega_{1})E^{\lambda}(\omega_{2})d_{ab}(\omega)d_{aa}(\omega-\omega_{1})\xi_{ab}^{\nu}\partial_{\lambda}f_{ab}2\pi\delta(\omega -\omega_{1} - \omega_{2})+ [(\nu, \omega_{1}) \leftrightarrow (\lambda, \omega_{2})],    
        \end{split},\\
        \begin{split}
            \rho_{EE, ab}^{(ee)} &= \dfrac{1}{2}\int \dfrac{d\omega_{1} d\omega_{2}}{(2\pi)^{2}}E^{\nu}(\omega_{1})E^{\lambda}(\omega_{2})\qty(d(\omega) \odot [r_{e}^{\nu}, d(\omega - \omega_{1})\odot [r_{e}^{\lambda}, \rho^{(0)}(\omega-\omega_{1}-\omega_{2})]])_{ab} + [(\nu, \omega_{1}) \leftrightarrow (\lambda, \omega_{2})] \\
        &= \dfrac{1}{2}\sum_{c}\int \dfrac{d\omega_{1} d\omega_{2}}{(2\pi)^{2}}E^{\nu}(\omega_{1})E^{\lambda}(\omega_{2})d_{ab}(\omega)\qty[d_{cb}(\omega -\omega_{1})\xi_{ac}^{\mu}\xi_{cb}^{\mu}f_{bc} - d_{ac}(\omega - \omega_{1})\xi_{cb}^{\mu}\xi_{ac}^{\nu}f_{ca}]2\pi\delta(\omega-\omega_{1}-\omega_{2}) \\
        & \hspace{12cm}+ [(\nu, \omega_{1})\leftrightarrow (\lambda, \omega_{2})].
        \end{split}
\end{align}
Due to the Fermi surface factor $\partial_{\nu} f$, $\rho^{(ii)}_{EE}, \rho^{(ei)}_{EE}$ vanishes in the limit of cold insulators. In the following calculation, we neglect these components and derive the shift and injection current formula. 

Using $\rho_{EE,ab}$, we can get the following second-order current response,
\begin{align}
        \begin{split}
            J^{\mu}_{EE}(\omega) &= \int \dfrac{d\vb*{k}}{(2\pi)^{d}}\sum_{abc}J_{ab}^{\mu}\rho_{EE, ba}(\omega) \\
        &\eqqcolon  \int\dfrac{d\omega_{1}d\omega_{2}}{(2\pi)^{2}}{\sigma}^{\mu;\nu\lambda}_{EE}(\omega, \omega_{1}, \omega_{2})E^{\nu}(\omega_{1})E^{\lambda}(\omega_{2})2\pi\delta(\omega - \omega_{1} - \omega_{2}).
        \end{split}
\end{align}
As we divide the SPDM into the four different contributions, $\sigma_{EE}^{\mu;\nu\lambda}(\omega, \omega_{1}, \omega_{2})$ is also decomposed into the corresponding the four components defined as,
\begin{align}
    \sigma_{EE}^{\mu;\nu\lambda} = \sigma_{EE, (ii)}^{\mu;\nu\lambda} + \sigma_{EE, (ie)}^{\mu;\nu\lambda} + \sigma_{EE, (ei)}^{\mu;\nu\lambda} + \sigma_{EE, (ee)}^{\mu;\nu\lambda}.
\end{align}
Photocurrent conductivity can be obtained by setting 
\begin{align}
        \omega =0 , \omega_{1} = -\Omega, \omega_{2} = \Omega.
\end{align}
In the following calculation, we focus on $ \sigma_{EE, (ie)}^{\mu;\nu\lambda},$ $\sigma_{EE, (ee)}^{\mu;\nu\lambda}$, from which we will derive shift current and injection current conductivity.
\subsection{Injection current}
Firstly, we focus on $\sigma_{EE, (ee)}^{\mu;\nu\lambda}$ with the diagonal component of the current matrix denoted as $\sigma_{EE, (ee;d)}^{\mu;\nu\lambda}$;
\begin{align}
        \begin{split}
            &\sigma^{\mu;\nu\lambda}_{EE,(ee;d)}(\omega,\omega_{1},\omega_{2}) \\
        &= \dfrac{1}{2}\int \dfrac{d\vb*{k}}{(2\pi)^{d}}\sum_{a\neq c}J_{aa}^{\mu}d_{aa}(\omega)\qty[d_{ca}(\omega-\omega_{1})\xi^{\nu}_{ac}\xi^{\lambda}_{ca}f_{ac} - d_{ac}(\omega-\omega_{1})\xi_{ca}^{\nu}\xi_{ac}^{\lambda}f_{cb}] + [(\nu, \omega_{1}) \leftrightarrow (\lambda, \omega_{2})] \\
        &= \dfrac{1}{2}\dfrac{1}{\omega+i\eta}\int \dfrac{d\vb*{k}}{(2\pi)^{d}}\sum_{a\neq c}(J_{aa}^{\mu} - J_{cc}^{\mu})\qty[d_{ca}(\omega-\omega_{1})\xi^{\nu}_{ac}\xi^{\lambda}_{ca}f_{ac}] + [(\nu, \omega_{1}) \leftrightarrow (\lambda, \omega_{2})] \\
        &= \dfrac{1}{2}\dfrac{1}{\omega+i\eta}\int \dfrac{d\vb*{k}}{(2\pi)^{d}}\sum_{a\neq c}(J_{aa}^{\mu} - J_{cc}^{\mu})\xi^{\nu}_{ac}\xi^{\lambda}_{ca}f_{ac}\qty[d_{ca}(\omega-\omega_{1}) + d_{ac}(\omega-\omega_{2})] \\
        &= \dfrac{1}{2}\dfrac{1}{\omega+i\eta}\int \dfrac{d\vb*{k}}{(2\pi)^{d}}\sum_{a\neq c}\Delta_{ac}^{\mu}\xi^{\nu}_{ac}\xi^{\lambda}_{ca}f_{ac}\qty[d_{ca}(\omega-\omega_{1}) + d_{ac}(\omega-\omega_{2})].
        \end{split}
\end{align}
Here we introduced velocity difference matrix $\Delta_{ac}^{\mu} = J_{aa}^{\mu} - J_{cc}^{\mu} = \partial_{\mu}\epsilon_{a} - \partial_{\mu}\epsilon_{b}$.
In the $\omega \to 0$ limit, the expression diverges due to the factor of $1/\omega+i\eta$. We can eliminate this unphysical divergence by considering the infinitesimal parameter $\eta$ as finite system relaxation $\gamma$. By using this method and relation \Eqref{Dirac_formula}, we can obtain the injection current formula as
\begin{align}
        \sigma_{EE,\text{inj}}^{\mu;\nu\lambda} 
        &= \dfrac{\pi}{\gamma}\int \dfrac{d\vb*{k}}{(2\pi)^{d}}\sum_{a\neq c}\Delta_{ac}^{\mu}\xi^{\nu}_{ac}\xi^{\lambda}_{ca}f_{ac}\delta(\Omega-\epsilon_{ca}).
\end{align}
Here we decompose the $\sigma_{EE,\text{inj}}^{\mu;\nu\lambda}$ into real and imaginary part as
\begin{align}
        \sigma_{EE, \text{Mnj}}^{\mu;\nu\lambda} &= \dfrac{\pi}{\gamma}\int \dfrac{d\vb*{k}}{(2\pi)^{d}}\sum_{a\neq c}\Delta_{ac}^{\mu}\operatorname{Re}\qty[\xi^{\nu}_{ac}\xi^{\lambda}_{ca}]f_{ac}\delta(\Omega-\epsilon_{ca}),\\
        \sigma_{EE, \text{Enj}}^{\mu;\nu\lambda} &= i\dfrac{\pi}{\gamma}\int \dfrac{d\vb*{k}}{(2\pi)^{d}}\sum_{a\neq c}\Delta_{ac}^{\mu}\operatorname{Im}\qty[\xi^{\nu}_{ac}\xi^{\lambda}_{ca}]f_{ac}\delta(\Omega-\epsilon_{ca}).
\end{align}
$\sigma_{\text{Mnj}}$ ($\sigma_{\text{Enj}}$) are called magnetic (electric) injection current, which is induced by linearly (circularly) polarized light. 
\subsection{Shift current}
Here we derive the shift current formula based on $\sigma_{EE, (ee)}^{\mu;\nu\lambda}$ and $\sigma_{EE, (ie)}^{\mu;\nu\lambda}$. First, we consider $\sigma_{EE, (ee)}^{\mu;\nu\lambda}$ with the off-diagonal component of the velocity operator denoted as $\sigma^{\mu;\nu\lambda}_{EE, (ee;o)}$;
\begin{align}
        \begin{split}
            &\sigma^{\mu;\nu\lambda}_{EE, (ee;o)}(0, -\Omega, \Omega) \\
        &= \dfrac{1}{2}\int \dfrac{d\vb*{k}}{(2\pi)^{d}}\sum_{a\neq b\neq c}J_{ab}^{\mu}d_{ba}(0)\qty[d_{ca}(\Omega)\xi^{\nu}_{bc}\xi^{\lambda}_{ca}f_{ac} - d_{bc}(\Omega)\xi_{ca}^{\nu}\xi_{bc}^{\lambda}f_{cb}] + [(\nu, -\Omega) \leftrightarrow (\lambda, \Omega)] \\
        &= \dfrac{1}{2}\int \dfrac{d\vb*{k}}{(2\pi)^{d}}\sum_{a\neq b\neq c}i\epsilon_{ab}\xi_{ab}^{\mu}d_{ba}(0)\qty[d_{ca}(\Omega)\xi^{\nu}_{bc}\xi^{\lambda}_{ca}f_{ac} - d_{bc}(\Omega)\xi_{ca}^{\nu}\xi_{bc}^{\lambda}f_{cb}] + [(\nu, -\Omega) \leftrightarrow (\lambda, \Omega)] \\
        &= \dfrac{1}{2}\int \dfrac{d\vb*{k}}{(2\pi)^{d}}\sum_{a\neq b \neq c}i\xi_{ab}^{\mu}\qty[d_{ca}(\Omega)\xi^{\nu}_{bc}\xi^{\lambda}_{ca}f_{ac} - d_{bc}(\Omega)\xi_{ca}^{\nu}\xi_{bc}^{\lambda}f_{cb}] + [(\nu, -\Omega) \leftrightarrow (\lambda, \Omega)] \\
        &=  \dfrac{1}{2}\int \dfrac{d\vb*{k}}{(2\pi)^{d}}\sum_{a\neq b \neq c}i(\xi_{ab}^{\mu}\xi_{bc}^{\nu} - \xi_{bc}\xi_{ab}^{\nu})\xi_{ca}^{\lambda}f_{ac}d_{ac}(\Omega) + [(\nu, -\Omega) \leftrightarrow (\lambda, \Omega)] \\
        &= \dfrac{1}{2}\int \dfrac{d\vb*{k}}{(2\pi)^{d}}\sum_{a\neq c}\qty(\qty[D^{\mu}\xi^{\nu}]_{ac}- \qty[D^{\nu}\xi^{\mu}]_{ac})\xi_{ca}^{\lambda}f_{ac}d_{ac}(\Omega)+ [(\nu, -\Omega) \leftrightarrow (\lambda, \Omega)].
        \end{split}
\end{align}
In the last line, we defined $U(1)$-covariant derivative 
\begin{align}
        \qty[D^{\mu}O]_{ab} = \pdv{O_{ab}}{k^{\mu}} - i\qty(\xi_{aa}^{\mu} - \xi_{bb}^{\mu})O_{ab},
\end{align}
and applied the following formula under the condition of $a \neq c$;
\begin{align}
        \qty[D^{\mu}\xi^{\nu}]_{ac}- \qty[D^{\nu}\xi^{\mu}]_{ac} = \sum_{b}i(\xi_{ab}^{\mu}\xi_{bc}^{\nu} - \xi_{bc}^{\mu}\xi_{ab}^{\nu}).
\end{align}
Next, we consider the $\sigma_{EE, (ie)}^{\mu;\nu\lambda}$ component;
\begin{align}
        \begin{split}
            &\sigma^{\mu;\nu\lambda}_{EE, (ie)}(0;-\Omega,\Omega) \\
        &= -\dfrac{i}{2}\int \dfrac{d\vb*{k}}{(2\pi)^{d}}\sum_{a\neq b}J_{ab}^{\mu}d_{ba}(\omega) \qty[\pdv{k^{\nu}}\qty(d_{ba}(\Omega)f_{ba}\xi_{ba}^{\lambda} - i \qty(\xi_{aa}^{\nu} - \xi_{bb}^{\mu})d_{ba}(\Omega)f_{ba}\xi_{ba}^{\lambda})] + [(\nu, -\Omega) \leftrightarrow (\lambda, \Omega)] \\
        &= -\dfrac{i}{2}\int \dfrac{d\vb*{k}}{(2\pi)^{d}}\sum_{a\neq b}i\epsilon_{ab}\xi_{ab}^{\mu}d_{ba}(\omega) \qty[\pdv{k^{\nu}}\qty(d_{ba}(\Omega)f_{ba}\xi_{ba}^{\lambda} - i \qty(\xi_{aa}^{\nu} - \xi_{bb}^{\mu})d_{ba}(\Omega)f_{ba}\xi_{ba}^{\lambda})]+ [(\nu, -\Omega) \leftrightarrow (\lambda, \Omega)]  \\
        &= \dfrac{1}{2}\int \dfrac{d\vb*{k}}{(2\pi)^{d}}\sum_{a\neq b}\xi_{ab}^{\mu} \qty[\pdv{k^{\nu}}\qty(d_{ba}(\Omega)f_{ba}\xi_{ba}^{\lambda}) - i \qty(\xi_{aa}^{\nu} - \xi_{bb}^{\mu})d_{ba}(\Omega)f_{ba}\xi^{\lambda}_{ba}] + [(\nu, -\Omega) \leftrightarrow (\lambda, \Omega)] \\
        &= -\dfrac{1}{2}\int \dfrac{d\vb*{k}}{(2\pi)^{d}}\sum_{a\neq b} \qty[D^{\nu}\xi^{\mu}]_{ab} d_{ba}(\Omega)f_{ba}\xi_{ba}^{\lambda}+ [(\nu, -\Omega) \leftrightarrow (\lambda, \Omega)].
        \end{split}
\end{align}
By summing up these contributions, we obtain
\begin{align}
        \begin{split}
            \sigma^{\mu;\nu\lambda}_{EE, (ee;o)+(ie)} &=  \sigma^{\mu;\nu\lambda}_{EE, (ee;o)}(0, -\Omega, \Omega) + \sigma^{\mu;\nu\lambda}_{EE, (ie)}(0;-\Omega,\Omega) \\
        &= \dfrac{1}{2}\int \dfrac{d\vb*{k}}{(2\pi)^{d}}\sum_{a\neq b}\qty[D^{\mu}\xi^{\nu}]_{ab}\xi_{ba}^{\lambda}f_{ab}d_{ba}(\Omega)+ [(\nu, -\Omega) \leftrightarrow (\lambda, \Omega)] \\
        &= \dfrac{1}{2}\int \dfrac{d\vb*{k}}{(2\pi)^{d}}\sum_{a\neq b}\qty[D^{\mu}\xi^{\nu}]_{ab}\xi_{ba}^{\lambda}f_{ab}d_{ba}(\Omega)+ \qty[D^{\mu}\xi^{\lambda}]_{ab}\xi_{ba}^{\nu}f_{ab}d_{ba}(-\Omega) \\
        &= \dfrac{1}{2}\int \dfrac{d\vb*{k}}{(2\pi)^{d}}\sum_{a\neq b}\qty[\qty[D^{\mu}\xi^{\nu}]_{ab}\xi_{ba}^{\lambda} + \qty[D^{\mu}\xi^{\lambda}]_{ba}\xi_{ab}^{\nu}]f_{ab}\mathcal{P}\dfrac{1}{\Omega-\epsilon_{ba}} \\
        &-\dfrac{i\pi}{2}\int \dfrac{d\vb*{k}}{(2\pi)^{d}}\sum_{a\neq b}\qty[\qty[D^{\mu}\xi^{\nu}]_{ab}\xi_{ba}^{\lambda} - \qty[D^{\mu}\xi^{\lambda}]_{ab}\xi_{ba}^{\nu}]f_{ab}\delta(\Omega-\epsilon_{ba}).
        \end{split}
\end{align}
Absorptive part $\delta(\Omega-\epsilon_{ba})$ corresponds to the shift current conductivity, and we divide the real and imaginary parts as 
\begin{align}
        \sigma_{EE,\text{shift}}^{\mu;\nu\lambda} = \dfrac{\pi}{2}\int \dfrac{d\vb*{k}}{(2\pi)^{d}}\sum_{a\neq b}\operatorname{Im}\qty[\qty[D^{\mu}\xi^{\nu}]_{ab}\xi_{ba}^{\lambda} - \qty[D^{\mu}\xi^{\lambda}]_{ba}\xi_{ab}^{\nu}]f_{ab}\delta(\Omega-\epsilon_{ba}), \\
        \sigma_{EE,\text{gyro}}^{\mu;\nu\lambda} = -\dfrac{i\pi}{2}\int \dfrac{d\vb*{k}}{(2\pi)^{d}}\sum_{a\neq b}\operatorname{Re}\qty[\qty[D^{\mu}\xi^{\nu}]_{ab}\xi_{ba}^{\lambda} - \qty[D^{\mu}\xi^{\lambda}]_{ba}\xi_{ab}^{\nu}]f_{ab}\delta(\Omega-\epsilon_{ba}).
\end{align}
$\sigma_{\text{shift}}$ ($\sigma_{\text{gyro}}$) are called shift (gyration) currents, which are induced by linearly (circularly) polarized light.

\subsection{Spin dynamics induced photocurrent}
Here, we derive the photocurrent formula induced by spin dynamics. Using the SPDM $\rho_{\mathrm{SS}}$, current response to the spin field can be expressed as
\begin{align}
        J^{\mu}_{\mathrm{SS}} &= \int \dfrac{d\vb*{k}}{(2\pi)^{d}}\sum_{abc}J_{ab}^{\mu}\rho_{\mathrm{S}\mathrm{S}, ba}^{(2)}(\omega)  \\
        &\eqqcolon \int\dfrac{d\omega_{1}d\omega_{2}}{(2\pi)^{2}}\sigma^{\mu;\nu\lambda}_{\mathrm{SS}}(\omega, \omega_{1}, \omega_{2})\Delta\mathrm{S}^{\nu}(\omega_{1})\Delta\mathrm{S}^{\lambda}(\omega_{2})2\pi\delta(\omega-\omega_{1}-\omega_{2})
\end{align}
In the previous subsection, we derive the photocurrent conductivity based on the perturbative expansion of the von Neumann equation and obtain the shift current and injection current formula. In the same manner, we can get the shift current and injection current-like formula, using $\rho_{\mathrm{SS}}$, just by replacing the Berry connection term $\xi$ related to the external light field appearing in the $\sigma_{EE}$ by spin operator $A$. As a result, we get the shift current and injection-like current formula induced by spin dynamics;
\begin{align}
        \sigma_{\mathrm{S}\mathrm{S},\text{shift}}^{\mu;\nu\lambda} &= J^{2}\dfrac{\pi}{2}\int \dfrac{d\vb*{k}}{(2\pi)^{d}}\sum_{a\neq b}\operatorname{Im}\qty[\qty[D^{\mu}A^{\nu}]_{ab}A_{ba}^{\lambda} - \qty[D^{\mu}A^{\lambda}]_{ba}A_{ab}^{\nu}]f_{ab}\delta(\Omega-\epsilon_{ba}), \\
        \sigma_{\mathrm{S}\mathrm{S},\text{gyro}}^{\mu;\nu\lambda} &= -J^{2}\dfrac{i\pi}{2}\int \dfrac{d\vb*{k}}{(2\pi)^{d}}\sum_{a\neq b}\operatorname{Re}\qty[\qty[D^{\mu}A^{\nu}]_{ab}A_{ba}^{\lambda} - \qty[D^{\mu}A^{\lambda}]_{ba}A_{ab}^{\nu}]f_{ab}\delta(\Omega-\epsilon_{ba}), \\
        \sigma_{\mathrm{S}\mathrm{S},\text{Mnj}}^{\mu;\nu\lambda} &= J^{2}\dfrac{\pi}{\gamma}\int \dfrac{d\vb*{k}}{(2\pi)^{d}}\sum_{a\neq c}\Delta_{ac}^{\mu}\operatorname{Re}\qty[A^{\nu}_{ac}A^{\lambda}_{ca}]f_{ac}\delta(\Omega-\epsilon_{ca}), \\
        \sigma_{\mathrm{S}\mathrm{S}, \text{Enj}}^{\mu;\nu\lambda} &= J^{2}\dfrac{i\pi}{\gamma}\int \dfrac{d\vb*{k}}{(2\pi)^{d}}\sum_{a\neq c}\Delta_{ac}^{\mu}\operatorname{Im}\qty[A^{\nu}_{ac}A^{\lambda}_{ca}]f_{ac}\delta(\Omega-\epsilon_{ca}).
\end{align}

\subsection{Interference of light field and spin dynamics}
Following the previous subsection, we derive the photocurrent formula generated by the interference of the light field and spin dynamics. Using $\rho_{E\mathrm{S}}$, photocurrent formula can be written as
\begin{align}
        J_{E\mathrm{S}}^{\mu}(\omega) &= \int \dfrac{d\vb*{k}}{(2\pi)^{d}}\sum_{abc}J_{ab}^{\mu}\rho_{E\mathrm{S},ba}^{(2)}(\omega) \\
        &\eqqcolon \int\dfrac{d\omega_{1}d\omega_{2}}{(2\pi)^{2}}\tilde{\sigma}_{E\mathrm{S}}^{\mu;\nu\lambda}(\omega, \omega_{1}, \omega_{2})E^{\nu}(\omega_{1})\Delta\mathrm{S}^{\lambda}(\omega_{2})2\pi\delta(\omega-\omega_{1}-\omega_{2}), 
\end{align}
where $\tilde{\sigma}_{E\mathrm{S}}^{\mu;\nu\lambda}$ can be classified into the following four component.
\begin{align}
        \sigma_{E\mathrm{S},\text{shift}}^{\mu;\nu\lambda} &= J\dfrac{\pi}{2}\int \dfrac{d\vb*{k}}{(2\pi)^{d}}\sum_{a\neq b}\operatorname{Im}\qty[\qty[D^{\mu}\xi^{\nu}]_{ab}A_{ba}^{\lambda} - \qty[D^{\mu}A^{\lambda}]_{ba}\xi_{ab}^{\nu}]f_{ab}\delta(\Omega-\epsilon_{ba}), \\
        \sigma_{E\mathrm{S},\text{gyro}}^{\mu;\nu\lambda} &= -J\dfrac{i\pi}{2}\int \dfrac{d\vb*{k}}{(2\pi)^{d}}\sum_{a\neq b}\operatorname{Re}\qty[\qty[D^{\mu}\xi^{\nu}]_{ab}A_{ba}^{\lambda} - \qty[D^{\mu}A^{\lambda}]_{ba}\xi_{ab}^{\nu}]f_{ab}\delta(\Omega-\epsilon_{ba}), \\
        \sigma_{E\mathrm{S},\text{Mnj}}^{\mu;\nu\lambda} &= J\dfrac{\pi}{\gamma}\int \dfrac{d\vb*{k}}{(2\pi)^{d}}\sum_{a\neq c}\Delta_{ac}^{\mu}\operatorname{Re}\qty[\xi^{\nu}_{ac}A^{\lambda}_{ca}]f_{ac}\delta(\Omega-\epsilon_{ca}),\\
        \sigma_{E\mathrm{S}, \text{Enj}}^{\mu;\nu\lambda} &= J\dfrac{i\pi}{\gamma}\int \dfrac{d\vb*{k}}{(2\pi)^{d}}\sum_{a\neq c}\Delta_{ac}^{\mu}\operatorname{Im}\qty[\xi^{\nu}_{ac}A^{\lambda}_{ca}]f_{ac}\delta(\Omega-\epsilon_{ca}).
\end{align}

\section{Symmetry analysis of the photocurrent}\label{sym_analysis_photocurrent}
In this section, we show the symmetry analysis of photocurrent in the presence of spin dynamics. Due to the $\theta 2_{x}$ symmetry of the system and phase matching between the different fields, the photocurrent generation is restricted. Here, we consider a one-dimensional system, applying the light field along the $z$ direction and observing the photocurrent along the $z$ direction, as in the case of the main text. In the following calculation, we assume the wave vector $\vb*{k}$ to be scaler.

Under $\theta 2_{x}$ operation, the wave vector along $z$ direction transforms as  $k \rightarrow k$. Therefore, Bloch states that momentum $k$ is related to those with the same momentum $k$. 
\begin{align}
        \theta 2_{x} \ket{u_{a}(k)} = \ket{{u_{\tilde{a}}}(k)}.
\end{align}
Since $(\theta2_{x})^{2} = 1$, there is no Kramers degeneracy. Therefore, by choosing the phase factor, we can assume 
\begin{align}
        \theta 2_{x} \ket{u_{a}(k)} = \ket{{u_{a}}(k)}. 
\end{align}
Due to the $\theta2_{x}$ symmetry, Berry connection $\xi_{ab}^{z}(k)$ satisfies the following relation
\begin{align}
        \begin{split}
            \xi_{ab}^{z}(k) &=  \matrixel{u_{a}(k)}{i\partial_{k}}{u_{b}(k)}\\
        &= \matrixel{u_{b}(k)}{(\theta2_{x})i\partial_{k}(\theta2_{x})^{-1}}{u_{a}(k)} \\
        &= \matrixel{u_{b}(k)}{-i\partial_{k}}{u_{a}(k)} \\
        &= -\xi_{ba}^{z}(k).
        \end{split}
\end{align}
Moreover, the $A^{\nu}$ under the $\theta 2_{x}$ symmetry satisfies the following relation.
\begin{align}
        \begin{split}
            A_{ab}^{\nu} &= \matrixel{u_{a}(k)}{A^{\nu}}{u_{b}(k)} \\
        &= \matrixel{u_{b}(k)}{(\theta 2_{x})(A^{\nu})^{\dagger}(\theta 2_{x})}{u_{a}(k)} \\
        &= \matrixel{u_{b}(k)}{(\theta 2_{x})(A^{\nu})(\theta 2_{x})}{u_{a}(k)} \\
        &= \sigma_{A^{\nu}}\matrixel{u_{b}(k)}{A^{\nu}}{u_{a}(k)} \\
        &= \sigma_{A^{\nu}}A_{ba}^{\nu}.
        \end{split}
\end{align}
Here $\sigma_{A^{\nu}}$ is the sign under $\theta 2_{x}$ operation. As for the spin operator coupled to the $\alpha$-mode, $\sigma_{A^{\nu}}$ satisfies the following relation.
\begin{align}
        \sigma_{\sigma_{x}\tau_{z}} = -1, 
        \sigma_{\sigma_{y}\tau_{0}} = 1, 
        \sigma_{\sigma_{z}\tau_{z}} = 1.\label{theta2x_symmetry}
\end{align}
\subsection{Light field induced photocurrent}
Drawing on the previous subsection, the photocurrent along the $z$ direction induced by the light field along the $z$ direction can be written as
\begin{align}
        J_{EE}^{z} = \int \dfrac{d\Omega}{2\pi}\sigma_{EE}^{z;zz}(0;-\Omega,\Omega)E^{z}(-\Omega)E^{z}(\Omega),\label{J_PVE_IPA}
\end{align} 
where $\sigma^{z;zz}$ can be classified into following four contributions,
\begin{align}
        \sigma_{EE, \text{shift}}^{z;zz} &= \dfrac{\pi}{2}\int \dfrac{dk}{2\pi}\sum_{a\neq b}\operatorname{Im}\qty[\qty[D^{z}\xi^{z}]_{ab}\xi_{ba}^{z} - \qty[D^{z}\xi^{z}]_{ba}\xi_{ab}^{z}]f_{ab}\delta(\Omega-\epsilon_{ab}), \\
        \sigma_{EE, \text{gyro}}^{z;zz} &= -\dfrac{i\pi}{2}\int \dfrac{dk}{2\pi}\sum_{a\neq b}\operatorname{Re}\qty[\qty[D^{z}\xi^{z}]_{ab}\xi_{ba}^{z} - \qty[D^{z}\xi^{z}]_{ba}\xi_{ab}^{z}]f_{ab}\delta(\Omega-\epsilon_{ab}),\\
        \sigma_{EE, \text{Mnj}}^{z;zz} &= \dfrac{\pi}{\gamma}\int \dfrac{dk}{2\pi}\sum_{a\neq b}\Delta_{ab}^{z}\operatorname{Re}\qty[\xi^{z}_{ab}\xi^{z}_{ba}]f_{ab}\delta(\Omega-\epsilon_{ba}),\\
        \sigma_{EE, \text{Enj}}^{z;zz} &= i\dfrac{\pi}{\gamma}\int \dfrac{dk}{2\pi}\sum_{a\neq b}\Delta_{ab}^{z}\operatorname{Im}\qty[\xi^{z}_{ab}\xi^{z}_{ba}]f_{ab}\delta(\Omega-\epsilon_{ba}).
\end{align}

In \Eqref{J_PVE_IPA}, since the left hand side is real, and $E^{z}(\Omega)\qty[E^{z}(\Omega)]^{\ast}$ is also real, $\sigma^{z;zz}_{EE}(\Omega)$ should be real. Therefore only $\sigma_{\text{shift}}^{z;zz}, \sigma_{\text{Mnj}}^{z;zz}$ can contribute to the photocurrent generation. Moreover due to the $\theta 2_{x}$ in the system, covariant derivative $\qty[D^{z}\xi^{z}]_{ab}\xi_{ba}^{z}$ satisfies following relation
\begin{align}
        \begin{split}
            \qty[D^{z}\xi^{z}]_{ab}\xi^{z}_{ba} &= \qty[\pdv{\xi^{z}_{ab}}{k^{z}} - i\qty(\xi^{z}_{aa} - \xi_{bb}^{z})\xi_{ab}^{z}]\xi_{ba}^{z} \\
        &=  -\qty[-\pdv{\xi^{z}_{ba}}{k} - i\qty(\xi^{z}_{aa} - \xi^{z}_{bb})\xi^{z}_{ba}]\xi^{z}_{ab} \\
        &= \qty[\pdv{\xi^{z}_{ba}}{k} - i\qty(\xi^{z}_{bb} - \xi^{z}_{aa})\xi^{z}_{ba}]\xi^{z}_{ab} \\
        &= \qty[D^{z}\xi^{z}]_{ba}\xi^{z}_{ab}\\
        &= \qty(\qty[D^{z}\xi^{z}]_{ab}\xi^{z}_{ba})^{\ast}.
        \end{split}
\end{align}
Therefore $\sigma_{\text{shift}}$ contribution vanishes under the $\theta 2_{x}$ symmetry, while $\sigma_{\text{Mnj}}$ remain intact. 

\subsection{Spin dynamics induced photocurrent}
Photocurrent induced solely by collective spin dynamics can be expressed as 
\begin{align}
        \begin{split}
            J^{\mu}_{SS} &= \int\dfrac{d\Omega}{2\pi}\sigma_{\mathrm{SS}}^{\mu;\nu\lambda}(0;-\Omega,\Omega)\Delta \mathrm{S}^{\nu}(-\Omega)\Delta \mathrm{S}^{\lambda}(\Omega) \\
        &= \int\dfrac{d\Omega}{2\pi}\sigma_{\mathrm{SS}}^{\mu;\nu\lambda}(0;-\Omega,\Omega)\qty[\Delta \mathrm{S}^{\nu}(\Omega)]^{\ast}\Delta \mathrm{S}^{\lambda}(\Omega). \label{spindynamics_photocurrent}
        \end{split}
\end{align}
Here $\sigma_{\mathrm{SS}}$ can be classified into the following four components
\begin{align}
         \sigma_{\mathrm{S}\mathrm{S},\text{shift}}^{\mu;\nu\lambda} &= J^{2}\dfrac{\pi}{2}\int \dfrac{d\vb*{k}}{(2\pi)^{d}}\sum_{a\neq b}\operatorname{Im}\qty[\qty[D^{\mu}A^{\nu}]_{ab}A_{ba}^{\lambda} - \qty[D^{\mu}A^{\lambda}]_{ba}A_{ab}^{\nu}]f_{ab}\delta(\Omega-\epsilon_{ba}), \\
        \sigma_{\mathrm{S}\mathrm{S},\text{gyro}}^{\mu;\nu\lambda} &= -J^{2}\dfrac{i\pi}{2}\int \dfrac{d\vb*{k}}{(2\pi)^{d}}\sum_{a\neq b}\operatorname{Re}\qty[\qty[D^{\mu}A^{\nu}]_{ab}A_{ba}^{\lambda} - \qty[D^{\mu}A^{\lambda}]_{ba}A_{ab}^{\nu}]f_{ab}\delta(\Omega-\epsilon_{ba}),\\
        \sigma_{\mathrm{S}\mathrm{S},\text{Mnj}}^{\mu;\nu\lambda} &= J^{2}\dfrac{\pi}{\gamma}\int \dfrac{d\vb*{k}}{(2\pi)^{d}}\sum_{a\neq c}\Delta_{ac}^{\mu}\operatorname{Re}\qty[A^{\nu}_{ac}A^{\lambda}_{ca}]f_{ac}\delta(\Omega-\epsilon_{ca}),\\
        \sigma_{\mathrm{S}\mathrm{S}, \text{Enj}}^{\mu;\nu\lambda} &= J^{2}\dfrac{i\pi}{\gamma}\int \dfrac{d\vb*{k}}{(2\pi)^{d}}\sum_{a\neq c}\Delta_{ac}^{\mu}\operatorname{Im}\qty[A^{\nu}_{ac}A^{\lambda}_{ca}]f_{ac}\delta(\Omega-\epsilon_{ca}).
\end{align}
By considering the $\theta2_{x}$ operation, the matrix elements in each photocurrent conductivity satisfy the following relations.
\begin{align}
        \operatorname{Re}\qty[\qty[D^{z}A^{\nu}]_{ab}A_{ba}^{\lambda} - \qty[D^{z}A^{\lambda}]_{ba}A_{ab}^{\nu}]
        &= \dfrac{1+\sigma_{A^{\nu}}\sigma_{A^{\lambda}}}{2}\qty(\qty[D^{z}A^{\nu}]_{ab}A_{ba}^{\lambda} - \qty[D^{z}A^{\lambda}]_{ba}A_{ab}^{\nu}),\\
        \operatorname{Im}\qty[\qty[D^{z}A^{\nu}]_{ab}A_{ba}^{\lambda} - \qty[D^{z}A^{\lambda}]_{ba}A_{ab}^{\nu}] &= \dfrac{1-\sigma_{A^{\nu}}\sigma_{A^{\lambda}}}{2i}\qty(\qty[D^{z}A^{\nu}]_{ab}A_{ba}^{\lambda} - \qty[D^{z}A^{\lambda}]_{ba}A_{ab}^{\nu}),\\
        \operatorname{Re}\qty[A_{ab}^{\nu}A_{ba}^{\lambda}] &= \dfrac{1+\sigma_{A^{\nu}}\sigma_{A^{\lambda}}}{2}A_{ab}^{\nu}A_{ba}^{\lambda},\\
        \operatorname{Im}\qty[A_{ab}^{\nu}A_{ba}^{\lambda}] &=  \dfrac{1-\sigma_{A^{\nu}}\sigma_{A^{\lambda}}}{2}A_{ab}^{\nu}A_{ba}^{\lambda}.
\end{align}

In the equation \Eqref{spindynamics_photocurrent}, when two fields $\mathrm{S}^{\nu}$ and $\mathrm{S}^{\lambda}$ are in-phase (out-of-phase), the quantity $\qty[\Delta \mathrm{S}^{\nu}(\Omega)]^{\ast}\Delta \mathrm{S}^{\lambda}(\Omega)$ becomes real (pure-imaginary). Consequently, the photocurrent induced by two fields that are in phase with each other can be characterized by the real part of $\sigma_{\mathrm{SS}}$, namely $\sigma_{\mathrm{SS, shift}}$ and $\sigma_{\mathrm{SS, Mnj}}$. Conversely, the photocurrent induced by two fields that are out-of-phase with each other can be characterized by the imaginary part of $\sigma_{\mathrm{SS}}$, namely $\sigma_{\mathrm{SS, gyro}}$ and $\sigma_{\mathrm{SS, Enj}}$.
In our calculation, we confirmed that $\mathrm{M}^{y}$ and $\mathrm{L}^{z}$ are in-phase each other, while $\mathrm{L}^{x}$ is not perfectly in or out of phase with $\mathrm{M}^{y}$ and $\mathrm{L}^{z}$. Combining this phase analysis with the $\theta 2_{x}$ symmetry constraints in \Eqref{theta2x_symmetry}, we can identify the photocurrent contribution as summarized in \tabref{photocurrent_classification}. Although we consider the photocurrent response to the linearly polarized light, $\sigma_{\mathrm{SS}}$ contains the $\sigma_{\text{Enj}}$ which is the counterpart of circularly polarized induced photocurrent. Moreover, the photocurrent induced by spin dynamics can contain the shift current contribution, namely $\sigma_{\text{shift}}$ and $\sigma_{\text{gyro}}$, which may survive in the disordered system \cite{Hatada2020}.

\subsection{Interference of light field and spin dynamics}
Photocurrent generated by the interference of light field and spin dynamics can be described as follows.
\begin{align}
        J^{z}_{E\mathrm{S}} = \int\dfrac{d\Omega}{2\pi}\sigma_{E\mathrm{S}}^{z;z\lambda}(0;-\Omega,\Omega)E^{z}(-\Omega)\Delta \mathrm{S}^{\lambda}(\Omega). \label{J_ES}
\end{align}
Here $\sigma_{E\mathrm{S}}^{z;z\lambda}$ have four contributions and are explicitly written as follows.
\begin{align}
        \sigma_{E\mathrm{S},\text{shift}}^{z;z\lambda} &= J\dfrac{\pi}{2}\int \dfrac{d\vb*{k}}{(2\pi)^{d}}\sum_{a\neq b}\operatorname{Im}\qty[\qty[D^{z}\xi^{z}]_{ab}A_{ba}^{\lambda} - \qty[D^{z}A^{\lambda}]_{ba}\xi_{ab}^{z}]f_{ab}\delta(\Omega-\epsilon_{ba}), \\
        \sigma_{E\mathrm{S},\text{gyro}}^{z;z\lambda} &= -J\dfrac{i\pi}{2}\int \dfrac{d\vb*{k}}{(2\pi)^{d}}\sum_{a\neq b}\operatorname{Re}\qty[\qty[D^{z}\xi^{z}]_{ab}A_{ba}^{\lambda} - \qty[D^{z}A^{\lambda}]_{ba}\xi_{ab}^{z}]f_{ab}\delta(\Omega-\epsilon_{ba}), \\
        \sigma_{E\mathrm{S},\text{Mnj}}^{z;z\lambda} &= J\dfrac{\pi}{\gamma}\int \dfrac{d\vb*{k}}{(2\pi)^{d}}\sum_{a\neq c}\Delta_{ac}^{z}\operatorname{Re}\qty[\xi^{z}_{ac}A^{\lambda}_{ca}]f_{ac}\delta(\Omega-\epsilon_{ca}),\\
        \sigma_{E\mathrm{S}, \text{Enj}}^{z;z\lambda} &= J\dfrac{i\pi}{\gamma}\int \dfrac{d\vb*{k}}{(2\pi)^{d}}\sum_{a\neq c}\Delta_{ac}^{z}\operatorname{Im}\qty[\xi^{z}_{ac}A^{\lambda}_{ca}]f_{ac}\delta(\Omega-\epsilon_{ca}).
\end{align}
Under the $\theta2_{x}$ operation, the matrix element related to photocurrent generation satisfies the following relations.
\begin{align}
        \operatorname{Im}\qty[\qty[D^{z}\xi^{\nu}]_{ab}A_{ba}^{\lambda} - \qty[D^{z}A^{\lambda}]_{ab}\xi_{ba}^{\lambda}]&=\dfrac{1+\sigma_{A^{\lambda}}}{2i}\qty[\qty[D^{z}\xi^{\nu}]_{ab}A_{ba}^{\lambda} - \qty[D^{z}A^{\lambda}]_{ab}\xi_{ba}^{\lambda}], \\
        \operatorname{Re}\qty[\qty[D^{z}\xi^{\nu}]_{ab}A_{ba}^{\lambda} - \qty[D^{z}A^{\lambda}]_{ab}\xi_{ba}^{\lambda}]&=\dfrac{1-\sigma_{A^{\lambda}}}{2}\qty[\qty[D^{z}\xi^{\nu}]_{ab}A_{ba}^{\lambda} - \qty[D^{z}A^{\lambda}]_{ab}\xi_{ba}^{\lambda}], \\
        \operatorname{Re}\qty[\xi_{ac}^{\nu}A_{ca}^{\lambda}] &= \dfrac{1-\sigma_{A}^{\lambda}}{2}\xi_{ac}^{\nu}A_{ca}^{\lambda},\\
        \operatorname{Im}\qty[\xi_{ac}^{\nu}A_{ca}^{\lambda}] &= \dfrac{1+\sigma_{A}^{\lambda}}{2}\xi_{ac}^{\nu}A_{ca}^{\lambda}.
\end{align}
Besides the $\theta 2_{x}$ symmetry restriction, phase degrees of freedom between the light field and fictitious spin field play an important role in determining the photocurrent contribution. 
Rewriting the \Eqref{J_ES} by using the electromagnetic susceptibility $\chi_{\mathrm{S}^{\lambda}E}$ to the light field, we get
\begin{align}
        \begin{split}
            J_{ES}^{z} &= \int \dfrac{d\Omega}{2\pi}\sigma_{E\mathrm{S}}^{z;z\lambda}(0;-\Omega,\Omega)\chi_{\mathrm{S}^{\lambda}E}(\omega)E^{z}(-\Omega)E^{z}(\Omega) \\
        &= \int \dfrac{d\Omega}{2\pi}\sigma_{E\mathrm{S}}^{z;z\lambda}(0;-\Omega,\Omega)\qty[\operatorname{Re}\chi_{\mathrm{S}^{\lambda}E^{z}}(\Omega) + i\operatorname{Im}\chi_{\mathrm{S}^{\lambda}E^{z}}(\Omega)]\qty[E^{z}(\Omega)]^{\ast}E^{z}(\Omega) \\
        &= \int\dfrac{d\Omega}{2\pi}\qty[\operatorname{Re}\sigma_{E\mathrm{S}}^{z;z\lambda}(0;-\Omega,\Omega)\operatorname{Re}\chi_{\mathrm{S}^{\lambda}E^{z}}(\Omega) - \operatorname{Im}\sigma_{E\mathrm{S}}^{z;z\lambda}(0;-\Omega,\Omega)\operatorname{Im}\chi_{\mathrm{S}^{\lambda}E^{z}}(\Omega)]\qty[E^{z}(\Omega)]^{\ast}E^{z}(\Omega).
        \end{split}
\end{align}
Here, we used the fact that the left-hand side is real, and $\qty[E^{z}(\Omega)]^{\ast}E^{z}(\Omega)$ is also real. 
By considering the symmetry constraints, the photocurrent generation by the interference of light field and spin dynamics are summarized in the table \ref{photocurrent_classification}. Unlike the case of IPA, the generation of various kinds of photocurrent can be allowed in the $\sigma_{\text{ES}}$. 
\clearpage

\twocolumngrid

\end{document}